\documentclass[prb,aps,twocolumn,superscriptaddress,showpacs]{revtex4}
\usepackage{epsfig}
\usepackage{amsmath}
\usepackage{amsfonts}
\usepackage{amssymb,mathrsfs}

\newcommand{\bs}[1]{\boldsymbol{#1}}

\begin{document}

\title{Coulomb drag in graphene: perturbation theory}

\author{B.N. Narozhny} \affiliation{Institut f\"ur Theorie der
  Kondensierten Materie and DFG Center for Functional Nanostructures,
  Karlsruher Institut f\"ur Technologie, 76128 Karlsruhe, Germany}

\author{M. Titov} \affiliation{School of Engineering \& Physical
  Sciences, Heriot-Watt University, Edinburgh EH14 4AS, UK}
\affiliation{Institut f\"ur Nanotechnologie,
  Karlsruhe Institute of Technology, D-76021 Karlsruhe, Germany}

\author{I.V. Gornyi} \affiliation{Institut f\"ur Nanotechnologie,
  Karlsruhe Institute of Technology, D-76021 Karlsruhe, Germany}
\affiliation{A.F. Ioffe Physico-Technical Institute, 194021
  St. Petersburg, Russia}

\author{P.M. Ostrovsky} \affiliation{Institut f\"ur Nanotechnologie,
  Karlsruhe Institute of Technology, D-76021 Karlsruhe, Germany}
\affiliation{L.D. Landau Institute for Theoretical Physics RAS, 119334
  Moscow, Russia}

\date{\today}

\begin{abstract}
 We study the effect of Coulomb drag between two closely positioned
 graphene monolayers. In the limit of weak electron-electron
 interaction and small inter-layer spacing ($\mu_{1(2)}, T\ll v/d$)
 the drag is described by a universal function of the chemical
 potentials of the layers $\mu_{1(2)}$ measured in the units of
 temperature $T$. When both layers are tuned close to the Dirac point,
 then the drag coefficient is proportional to the product of the
 chemical potentials $\rho_D\propto\mu_1\mu_2$. In the opposite limit
 of low temperature the drag is inversely proportional to both
 chemical potentials $\rho_D\propto T^2/(\mu_1\mu_2)$. In the mixed
 case where the chemical potentials of the two layers belong to the
 opposite limits $\mu_1\ll T\ll\mu_2$ we find $\rho_D\propto
 \mu_1/\mu_2$. For stronger interaction and larger values of $d$ the
 drag coefficient acquires logarithmic corrections and can no longer
 be described by a power law. Further logarithmic corrections are due
 to the energy dependence of the impurity scattering time in graphene
 (for $\mu_{1(2)}\gg T$ these are small and may be neglected).  In the
 case of strongly doped (or gated) graphene $\mu_{1(2)}\gg v/d\gg T$
 the drag coefficient acquires additional dependence on the
 inter-layer spacing and we recover the usual Fermi-liquid result if
 the screening length is smaller than $d$.
\end{abstract}

%(when any of the layers is precisely at the Dirac point, then the drag
%vanishes due to electron-hole symmetry).


\maketitle

%\section{Introduction}

Transport measurements are conceptually simplest and by far the most
common experimental tools for studying inner workings of solids.
Within linear response, the outcome of such measurements is determined
by the properties of the unperturbed system, which are often the
object of study. In a typical experiment a current is driven through a
conductor and the voltage drop along the conductor is measured. In
most conventional conductors at low temperatures the resulting
resistance is mostly determined by disorder (which is always present
in any sample), while interactions between charge carriers lead to
corrections that affect the temperature dependence \cite{aar}.

Consider now a drag measurement in a bi-layer system consisting of two
closely spaced but electrically isolated conductors\cite{roj}. Passing
a current $I_1$ through one of these conductors (``the active layer'')
is known to induce a voltage $V_2$ in the other conductor (``the
passive layer''). The ratio of this voltage to the driving current
$\rho_D=-V_2/I_1$ (the transresistivity or the drag coefficient) is a
measure of inter-layer interaction. At low enough temperatures the
drag is dominated by the direct Coulomb interaction between charge
carriers in both layers \cite{roj}.

The physics of the Coulomb drag is well understood if both layers are
in the Fermi liquid state \cite{kor,fl2}. The current in the passive
layer is created by exciting electron-hole pairs (each pair consisting
of an occupied state above the Fermi surface and an empty state below)
in a state characterized by finite momentum. The momentum comes from
the electron-hole excitations in the active layer created by the
driving current. The momentum transfer is due to the inter-layer
Coulomb interaction. Therefore it follows from the usual phase-space
considerations that the drag coefficient is proportional to the square
of the temperature $\rho_D\propto T^2$. Remarkably, this simple
argument is sufficient to describe the observed temperature dependence
of $\rho_D$ (deviations from the quadratic dependence are due to the
effect of phonons) \cite{roj}.

The phase-space argument however does not describe the physics of the
effect completely. Indeed, in the passive layer the momentum is
transferred equally to electrons and holes so that the resulting state
can carry current only in the case of electron-hole asymmetry.
Likewise, this asymmetry is necessary for the current-carrying state
in the active layer to be characterized by non-zero total momentum.
In the Fermi-liquid theory the electron-hole asymmetry can be
expressed \cite{dsm} (assuming either a constant impurity scattering
time or diffusive transport \cite{fn2}) as a derivative of the
single-layer conductivity $\sigma_{1(2)}$ with respect to the chemical
potential. In conventional semiconductors \cite{kor} the asymmetry
appears due to curvature of the conduction band spectrum (leading to
the energy dependence of the density of states (DoS) and/or diffusion
coefficient): $\partial\sigma_i/\partial\mu_i\approx\sigma_i/\mu_i$.

Theoretical calculations \cite{roj,kor} typically focus on the drag
conductivity $\sigma_D$. The experimentally measurable drag
coefficient $\rho_D$ is then obtained by inverting the $2\times 2$
conductivity matrix. To the lowest order in the inter-layer
interaction \cite{kor} (assuming $\sigma_D\ll\sigma_{1(2)}$) one
obtains
\begin{equation}
\label{d1}
%\rho_D = \frac{\sigma_D}{\sigma_1\sigma_2}.
\rho_D = \sigma_D/(\sigma_1\sigma_2).
\end{equation}
Combining the above arguments (and assuming that the single-layer
conductivities $\sigma_i$ are given by the Drude formula) we arrive at 
the Fermi-liquid result \cite{kor}
\begin{equation}
\label{d2}
\rho_D = \frac{\hbar}{e^2} \frac{T^2}{\mu_1\mu_2} A_{12},
\end{equation}
where $A_{12}$ depends on the matrix elements of the inter-layer
interaction, the Fermi momenta of the two layers, and the inter-layer
spacing $d$ (in the diffusive regime, where the mean-free path
$\ell\ll d$, $A_{12}$ contains an additional logarithmic dependence
\cite{kor}). The precise form of $A_{12}$ is well known and can be
obtained by means of either the diagrammatic formalism \cite{kor} or
the kinetic equation \cite{jho,fln}.

Recently drag measurements were performed in a system of two parallel
graphene sheets \cite{tut}. It was shown that this system offers much
greater flexibility compared to prior experiments in semiconductor
heterostructures \cite{ex1}. The drag coefficient depends on
the following parameters: (i) temperature $T$, (ii) chemical
potentials of the layers $\mu_i$, (iii) inter-layer spacing $d$, (iv)
mean-free path $\ell_i$, and (v) the interaction strength. Earlier
experiments\cite{ex1} were performed on samples with large inter-layer
spacing, such that $\mu\gg v/d$ (where $v$ is the Fermi velocity). In
contrast, the graphene-based system allows one to scan a wide range of
chemical potentials (by electrostatically controlling carrier density)
from the Fermi-liquid regime with $\mu_i\gg v/d, T$ to the Dirac point
$\mu=0$ where the drag vanishes due to electron-hole symmetry.

Most experiments\cite{roj,ex1} (including that of
Ref.~\onlinecite{tut}) are performed in the ballistic regime
$\ell_i\gg d$, where $\rho_D$ does not explicitly depend on disorder,
albeit the conductivities $\sigma_D$ and $\sigma_i$ do. The
graphene-based sample of Ref.~\onlinecite{tut} is characterized by
much smaller inter-layer spacing in comparison to previous experiments
\cite{roj}, with all data taken at
\begin{equation}
\label{td}
%T\ll \frac{v}{d}.
T < v/d.
\end{equation}
Therefore in this paper we do not consider temperatures larger than
the inverse inter-layer spacing, even though our approach remains
valid for $T\gtrsim v/d$.

In this paper we address the problem of the Coulomb drag in graphene
by means of the perturbation theory. In the leading order, i.e. in
the limit of weak interaction
\begin{equation}
\label{a}
%\alpha = \frac{e^2}{v} \ll 1,
\alpha = e^2/v \ll 1,
\end{equation}
the drag conductivity $\sigma_D$ is described by the standard
(Aslamazov-Larkin-type, see Fig.~\ref{diagram}) diagram
\cite{kor}. Unlike usual metals, the single-layer conductivity in
graphene comprises two competing contributions: one due to disorder
and one due to electron-electron interaction \cite{kas,kin,int,po2}.
We assume that the dominant scattering mechanism is due to
disorder. Hence throughout the paper we assume
\begin{equation}
\label{att}
\alpha^2 T \tau \ll 1 \ll T\tau,
\end{equation}
where $\tau$ is the impurity scattering time (in the case of
energy-dependent impurity scattering time this parameter should be
understood as $\tau({\rm max}[\mu,T])$, see Sec.~\ref{mft}). The
latter inequality ensures that the system is in the ballistic
regime. Under this condition \cite{po2} the single-layer
conductivities $\sigma_{1(2)}$ are again given by the Drude-like
formula with the impurity scattering time. As we will show,
$\sigma_D/\sigma_{1(2)}\lesssim\alpha^2T\tau\ll 1$, which allows us to
evaluate the drag coefficient $\rho_D$ using Eq.~(\ref{d1}). Our
theory can be further extended to the case of stronger interaction or
vanishing disorder, where the condition (\ref{att}) can be
lifted. However it is more convenient to perform such calculations
within the framework of the kinetic equation \cite{kas,kin}. The
results of that work are reported elsewhere
\cite{us2}.

\begin{table}
\caption{Asymptotic expressions for the drag coefficient (\ref{rdpt})
  in the limit of weak inter-layer interaction and small inter-layer
  spacing $\mu_{1(2)}, T\ll v/d$.}
\begin{ruledtabular}
\begin{tabular}{ccc}
parameter region & drag coefficient &\\
\hline\noalign{\smallskip}
${\mu_1,\mu_2\ll T}$ & 
$\rho_D\approx 1.41 \; \alpha^2 \displaystyle{\frac{\hbar}{e^2} 
\frac{\mu_1\mu_2}{T^2}}$ & Eq.(\ref{rd-12s})\\\noalign{\smallskip}
$\mu_1\ll T\ll \mu_2$ & 
$\rho_D\approx 5.8 \; \alpha^2 \displaystyle{\frac{\hbar}{e^2} 
\frac{\mu_1}{\mu_2}}$ & Eq.(\ref{rd-12})\\\noalign{\smallskip}
$T\ll \mu_1< \mu_2$ & 
$\rho_D\approx \alpha^2 \displaystyle{\frac{\hbar}{e^2} \frac{8\pi^2}{3}
\frac{T^2}{\mu_1\mu_2} \ln\frac{\mu_1}{T}}$ & Eq.(\ref{rdlt-2})\\
\end{tabular}
\end{ruledtabular}
\label{table}
\end{table}

In the limit of small inter-layer spacing $\mu_{1(2)}, T\ll v/d$ the
resulting drag coefficient is the ``universal'' function 
\begin{equation}
\label{rdpt}
\rho_D = \alpha^2\frac{\hbar}{e^2} \; r_0(\mu_1/T, \mu_2/T).
\end{equation}
The function $r_0(x_1, x_2)$ is characterized by the limiting
cases $\mu_i\ll T$ and $\mu_i\gg T$, 
%where analytical results may be obtained, 
see Table~\ref{table}. In the vicinity of the Dirac point
$\rho_D\propto\mu_1\mu_2/T^2$, while in the opposite limit
$\rho_D\propto T^2/(\mu_1\mu_2)$ (with additional logarithmic factors,
see Sec.~\ref{ar2}). In the ``mixed'' case $\mu_1\ll T\ll\mu_2$ we
find $\rho_D\propto\mu_1/\mu_2$. This regime is apparently realized in
the experiment of Ref.~\onlinecite{tut} when the bottom layer is in
proximity of the Dirac point.

For relatively large inter-layer spacing, $\mu_i\gg v/d$ (and thus
$\mu_i\gg T$), a new regime appears, where the drag coefficient
acquires additional dependence on the inter-layer spacing. Here the
assumption (\ref{a}) may be relaxed as the perturbation theory
remains valid also for intermediate values of $\alpha$ (see
Sec.~\ref{fd} where we assume $\mu_1=\mu_2$). Still, for $\mu\ll T$ we
find $\rho_D\propto\mu^2/T^2$, at $\mu\sim T$ the drag coefficient
reaches its maximum, and further decays for $\mu\gg T$. The latter
regime is described by a long crossover from the above logarithmic
behavior to the Fermi-liquid result which is only achieved in the
limit of the small screening length $\varkappa d \gg 1$. Thus
$\rho_D(\mu\gtrsim v/d)$ cannot be described by a single power law.

The remainder of this paper is organized as follows. In Sec.~\ref{pt}
we present the perturbative calculation that gives the drag
resistivity in the limit $\alpha\rightarrow 0$ and $d\rightarrow
0$. While the resulting expression can only be evaluated numerically,
we analyze all interesting limits analytically for the case of two
identical layers in Sec.~\ref{ar} and also for the experimentally
relevant case where the two layers are characterized by different
chemical potentials, see Sec.~\ref{ar2}. In Sec.~\ref{al} we discuss
drag at intermediate values of $\alpha$ and $d$ as well as the role of
energy-dependent scattering time. The paper is concluded by a brief
summary and a discussion of the experimental relevance of our
results. Throughout the paper we use the units with $\hbar=1$ and only
restore the Planck's constant in the results for $\rho_D$. Technical
details are relegated to Appendices.

\section{Perturbative calculation of the Coulomb drag}
\label{pt}

Consider the limit of weak interaction $\alpha\rightarrow 0$. In this
case the drag conductivity $\sigma_D$ can be calculated with the help
of the lowest-order diagram\cite{kor}
\begin{equation}
\label{sd}
\sigma^{\alpha\beta}_D = \frac{1}{16\pi T}
\sum_{\bs{q}}
\int \frac{d\omega}{\sinh^2\frac{\omega}{2T}}
\Gamma_1^\beta(\omega, \bs{q}) 
\Gamma_2^\alpha(\omega, \bs{q}) 
| {\cal D}^R_{12} |^2,
\end{equation}
where the subscripts $1$ and $2$ refer to the active and passive
layers respectively, ${\cal D}^R_{12}$ is the retarded propagator of
the inter-layer interaction, and $\Gamma_a^\alpha(\omega, \bs{q})$ is
the non-linear susceptibility (or the rectification coefficient).
%,see Section~\ref{gm}; details are relegated to Appendix~\ref{gamma}).

{
\begin{figure}
\begin{center}
\epsfig{file=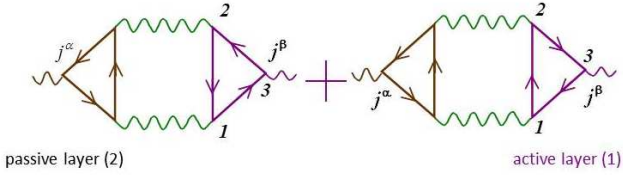,width=8.5cm}
\end{center}
\caption{[Color online] The lowest-order diagram for drag
  conductivity. Solid lines represent the exact single-electrons
  Green's functions in the presence of disorder. Spatial coordinates
  are labeled by the numbers shown at the corners of the triangles and
  correspond to the subscripts in Eq.~(\ref{g}). Wavy lines represent
  the inter-layer interaction.}
\label{diagram}
\end{figure}
}

\subsection{Inter-layer interaction for small $\alpha$ 
and $d$}

The bare Coulomb potential has the usual form
\begin{equation}
\label{bc}
V_{11} = V_{22} = \frac{2\pi e^2}{q}; \quad
V_{12} = \frac{2\pi e^2}{q} \; e^{-qd}.
\end{equation}
The dynamically screened (within RPA) inter-layer propagator can
be written as \cite{ady}
\begin{equation}
\label{d12}
{\cal D}^R_{12} = - \frac{1}
{\Pi^R_{1} \Pi^R_{2} \frac{4\pi e^2}{q} \sinh qd +
  \left(\frac{q}{2\pi e^2} + \Pi^R_{1} + \Pi^R_{2}\right) e^{qd}},
\end{equation}
where $\Pi^R_i$ is the single-layer retarded polarization operator.
For $\alpha\ll 1$ screening is ineffective and the
inter-layer interaction is essentially unscreened. Moreover, in the
limit $d\rightarrow 0$ we may disregard the exponential and write the
interaction propagator as
\begin{equation}
\label{dpt}
{\cal D}^R_{12} = - \frac{2\pi e^2}{q}.
\end{equation}
As we will show below, the non-linear susceptibility ${\bf \Gamma}$
decays exponentially for $q\gg{\rm max}(T, \mu)$. Therefore, in view of
Eq.~(\ref{td}) the limit $d\rightarrow 0$ is equivalent to the
condition $\mu\ll v/d$ (the case $\mu_i>v/d$ is discussed
in Sec.~\ref{fd}).

\subsection{Non-linear susceptibility in graphene}
\label{gm}

The non-linear susceptibility of electrons
$\Gamma_{ij}^\alpha(\omega)$ is a response function relating a voltage
$V(r_i)e^{i\omega t}$ to a $dc$ current it induces by the quadratic
response:
\begin{equation}
{\bf I}=\int d\bs{r}_1\int d\bs{r}_2
\bs{\Gamma}(\omega; \bs{r}_1,\bs{r}_2 ) 
V(\bs{r}_1)V(\bs{r}_2),
\label{gamma-el}
\end {equation}
\noindent
with ${\bf I}$ being the induced $dc$ current. From gauge invariance
it follows that
\[
\int d\bs{r}_1\bs{\Gamma}(\omega; \bs{r}_1,\bs{r}_2 ) = 
\int d\bs{r}_2\bs{\Gamma}(\omega; \bs{r}_1,\bs{r}_2 ) = 0.
\]

In terms of exact Green's functions of a disordered conductor
the non-linear susceptibility can be written as \cite{kor,gam}
\begin{eqnarray}
&&
{\bf \Gamma} = 
\int\frac{d\epsilon}{2\pi}
\Bigg[
      \left( \tanh\frac{\epsilon-\mu}{2T}
             - \tanh\frac{\epsilon+\omega-\mu}{2T}
      \right)
      \bs{\gamma}_{12}(\epsilon; \omega)
   %   \mbox{\boldmath$\gamma$}(\epsilon; \omega)
\nonumber\\
&&
\nonumber\\
&&
\quad
      +
      \left( \tanh\frac{\epsilon-\mu}{2T}
             - \tanh\frac{\epsilon-\omega-\mu}{2T}
      \right)
      \bs{\gamma}_{21}(\epsilon; -\omega)
\Bigg],
\label{nls}
\end{eqnarray}
\noindent
where the arguments of $\bs{\Gamma}$ are suppressed for brevity, the
subscripts $1$ and $2$ refer to the spatial coordinates $\bs{r}_1$ and
$\bs{r}_2$, respectively, and the triangular vertex $\bs{\gamma}$ is
given by
\begin{equation}
\label{g}
\bs{\gamma}_{12}(\epsilon; \omega) = 
\Big[
      G^R_{12}(\epsilon + \omega) - G^A_{12}(\epsilon + \omega)
\Big]
G^{R}_{23}\left(\epsilon\right) 
{\bf \hat J}_3 
G^{A}_{31}\left(\epsilon\right).
\end{equation}
Here ${\bf \hat J}_3$ is the current operator and integration over the
coordinate $\bs{r}_3$ is assumed (see Fig.~\ref{diagram}).

The traditional approach calls for averaging the vertex $\bs{\gamma}$
over disorder. Within the Fermi-liquid theory the averaged
$\bs{\gamma}$ does not depend on $\epsilon$ \cite{aar,kor}. This
result is based on the usual approximation that the most important
contribution comes from electrons near the Fermi surface. In contrast,
in graphene-based systems the regime where the chemical potential is
smaller that temperature $\mu_i\ll T$ is accessible. Then the vertex
$\bs{\gamma}$ retains its dependence on $\epsilon$ and should be
evaluated with care \cite{err,me1,dsa}. Details of the calculation
are provided in Appendices~\ref{gamma} and \ref{kin_ur}. The result is
\begin{eqnarray}
\label{gamma0}
&&
\bs{\gamma}(\epsilon, \omega, q) = -2\bs{q} ev\tau(\epsilon)
\frac{\omega^2+2\epsilon\omega-v^2q^2}{\epsilon v^2q^2}
\\
&&
\nonumber\\
&&
\quad\quad\quad\quad
\times
\sqrt{\frac{\left[2\epsilon+\omega\right]^2-v^2q^2}{v^2q^2-\omega^2}}
\;
{\rm sgn}(\epsilon+\omega) \theta_0(\epsilon, \omega, q),
\nonumber
\end{eqnarray}
where the function $\theta_0(\epsilon, \omega, q)$ ensures that the
expression under the square root in Eq.~(\ref{gamma0}) is always
positive (the minus signs in $\theta_0(\epsilon, \omega, q)$ reflect
the contribution of both electrons and holes and appear after summing
over all branches of the Dirac spectrum in graphene)
\begin{widetext}
\begin{eqnarray}
\label{theta0}
\theta_0(\epsilon, \omega, q) =
\theta(vq - |2\epsilon + \omega|)
\left[
\theta(-\omega-vq)-
\theta(\omega-vq)
\right]
+
\theta(vq-|\omega|)
\left[
\theta(2\epsilon + \omega - vq) -
\theta(- vq - 2\epsilon - \omega) \right].
\end{eqnarray}
Approximating \cite{fn3} the impurity scattering time by a constant
$\tau$ (the effect of energy-dependent $\tau(\epsilon)$ is discussed
below in Sec.~\ref{mft}), we can express the non-linear susceptibility
in terms of dimensionless variables
\[
W = \frac{\omega}{2T}, \quad Q = \frac{vq}{2T},
\]
and find the following form
%\begin{widetext}
\begin{subequations}
\label{gg}
\begin{equation}
\label{gg0}
\bs{\Gamma}(\omega, \bs{q}) = 
-2\frac{e\tau}{\pi} \; \bs{q} \; g\left(W, Q; \frac{\mu}{T}\right), 
\end{equation}
\begin{eqnarray}
\label{gg1}
\label{gg2}
g\left(W,Q; x\right) = 
\left\{
\begin{matrix}
\sqrt{\frac{W^2}{Q^2}-1}\int\limits_0^1 dz 
\displaystyle\frac{z\sqrt{1-z^2}}{z^2-W^2/Q^2}
\; I_2(z; W, Q; x), & |W|>Q \cr
\cr
-\sqrt{1-\frac{W^2}{Q^2}}\int\limits^\infty_1 dz \;
\displaystyle\frac{z\sqrt{z^2-1}}{z^2-W^2/Q^2} 
\; I(z; W, Q; x), & |W|<Q
\end{matrix}
\right.
\end{eqnarray}
\begin{eqnarray}
\label{i2}
I(z; W, Q; x) =
\tanh\frac{zQ+W+x}{2} -
\tanh\frac{zQ+W-x}{2} +
\tanh\frac{zQ-W-x}{2} -
\tanh\frac{zQ-W+x}{2}.
\end{eqnarray}
\end{subequations}
\end{widetext}

From Eq.~(\ref{i2}) it is clear that at the Dirac point the non-linear
susceptibility vanishes:
\begin{equation}
\label{gd}
I(x=0)=0 \; \Rightarrow \; \bs{\Gamma}(\mu=0) = 0.
\end{equation}
Thus there is no drag at the Dirac point (physically, due to
electron-hole symmetry).

\subsection{Perturbative results for the Coulomb drag in graphene}
\label{cdpt}

\subsubsection{Drag conductivity in graphene}

Using Eqs.~(\ref{dpt}) and (\ref{gg}) we can find the drag
conductivity in graphene (\ref{sd}). As usual for an isotropic system
in the absence of external magnetic fields $\sigma_D^{\alpha\beta} =
\delta ^{\alpha\beta} \sigma_D$.  Now, in the limit $\alpha\rightarrow
0$ and $d\rightarrow 0$ we find
\begin{subequations}
\label{r}
\begin{equation}
\label{sd-pt}
\sigma_D = \alpha^2 e^2 T^2 \tau^2 
f_0\left(\frac{\mu_1}{T}, \frac{\mu_2}{T}\right),
\end{equation}
where the dimensionless function $f_0(x_1, x_2)$ is defined as
\begin{equation}
\label{f0-2}
f_0(x_1,x_2) = \frac{4}{\pi^2} \int\limits_0^\infty QdQ
\int\limits_0^\infty \frac{dW}{\sinh^2 W} \;
g(x_1) g(x_2).
\end{equation}
\end{subequations}
where we have suppressed the arguments of the function $g\left(W,Q;
x\right)$ for brevity.

In general the function $f_0(x_1,x_2)$ has to be computed numerically.
Below, we evaluate this function analytically in the physically
interesting limiting cases of large $x$ (the low temperature regime)
and small $x$ (the vicinity of the Dirac point).

\subsubsection{Single-layer conductivity in graphene}

In order to find the drag coefficient $\rho_D$ one needs to know the
single-layer conductivity $\sigma_i$. Under our assumptions the
single-layer conductivity is completely determined by the weak
impurity scattering and can be written in the form
\begin{equation}
\label{s0}
\sigma_0 = e^2 T\tau h_0\left(\frac{\mu}{T}\right),
%\sigma_0 = e^2 T\tau \; h_0\left(\mu/T\right),
\end{equation}
where
\begin{equation}
\label{h0}
h_0(x) = \frac{2}{\pi}
\int\limits_{-\infty}^\infty dz 
\frac{|z|}{\cosh^2\left(z+\frac{x}{2}\right)}
=\frac{2}{\pi}\left\{
\begin{matrix}
x, & x\gg 1,\cr
2\ln 2, & x\ll 1.
\end{matrix}
\right.
\end{equation}

\subsubsection{The drag coefficient}

Finally, using Eqs.~(\ref{d1}), (\ref{sd-pt}), and (\ref{s0})
we find that the drag coefficient can indeed be expressed in the 
form (\ref{rdpt}), with the dimensionless function $r_0(x_1, x_2)$
defined as
\begin{equation}
\label{r0-1}
r_0(x_1, x_2) = \frac{f_0(x_1, x_2)}{h_0(x_1)h_0(x_2)}
\end{equation}
In the case of two identical layers this function depends on one
variable only
\[
r_0(x) = \frac{f_0(x, x)}{h_0^2(x)},
\]
and is shown in Fig.~\ref{r0}.

\newpage

The expressions (\ref{rdpt}) and (\ref{r0-1}) give the perturbative
result for the drag coefficient in graphene-based bi-layer systems in
terms of the dimensionless functions $h_0(x)$ and $f_0(x_1, x_2)$.
The applicability of this ``universal'' result and its experimental
relevance is discussed in Sec.~\ref{al}.

{
\begin{figure}
\begin{center}
\epsfig{file=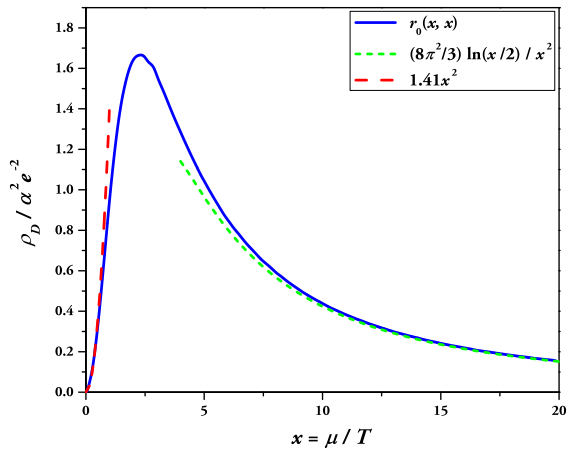,width=7.5cm}
\end{center}
\caption{[Color online] The dimensionless drag coefficient $r_0(x, x)$
  (two identical layers). The solid line shows the result of numerical
  evaluation using Eqs.~(\ref{r0-1}), (\ref{h0}), (\ref{f0-2}), and
  (\ref{gg}). The red dashed line shows the asymptotic behavior
  (\ref{rddp}) in the vicinity of the Dirac point. The green
  short-dashed line shows the asymptotic behavior (\ref{rdlt}) at
  small temperatures.}
\label{r0}
\end{figure}
}

\section{Asymptotic behavior of the drag between two identical layers}
\label{ar}

Consider now the case where the two layers in the drag experiment are
identical, i.e. are kept at the same temperature and chemical
potential. This case has not been yet realized experimentally in
graphene-based systems, but has a long history of theoretical research
\cite{roj,kor}.

\subsection{Vicinity of the Dirac point}

We start with situation where the graphene sheets are tuned close to
the Dirac point, i.e.
\[
\mu \ll T.
\]
Given that at the Dirac point the nonlinear susceptibility vanishes
[see Eq.~(\ref{gd})], we can expand it to the lowest order in the small
parameter $\mu/T$. Expanding first the expression (\ref{i2}) we find
\begin{equation}
\label{i2-3}
I(x\rightarrow 0) \approx - 4x
\frac{\sinh W \sinh zQ}
{\left(\cosh zQ + \cosh W \right)^2}.
\end{equation}
This corresponds to the quadratic expansion of Eq.~(\ref{f0-2})
\begin{equation}
\label{f0_dp}
f_0(x,x) \approx {\cal N}_1 x^2.
\end{equation}
The numerical coefficient ${\cal N}_1$ can now be determined by using
the approximation (\ref{i2-3}) in Eqs.~(\ref{gg1}) and (\ref{gg2}) and
then evaluating the integral in Eq.~(\ref{f0-2}). The result is
\begin{equation}
\label{n1}
{\cal N}_1 \approx 1.1 .
\end{equation}
Therefore the drag conductivity in the vicinity of the Dirac point is
independent of temperature and is proportional to the square of the
chemical potential
\begin{equation}
\label{sd-pt-l1}
\sigma_D(\mu\ll T) \approx \alpha^2 e^2 \mu^2 \tau^2 {\cal N}_1.
\end{equation}
In this case the drag conductivity is independent of $T$.

The single-layer conductivity can be found by
expanding the function $h_0(x)$. Thus in the vicinity of
the Dirac point the conductivity of a single graphene sheet
due to impurity scattering [see Eq.~(\ref{att})] is \cite{po2}
\begin{equation}
\label{s0dp}
\sigma_0\approx\frac{4\ln 2}{\pi} \; e^2 T\tau.
\end{equation}
Therefore the drag coefficient is
\begin{equation}
\label{rddp}
\rho_D(\mu\ll T) \approx \alpha^2 \frac{\hbar}{e^2}
\frac{\pi^2{\cal N}_1}{16\ln^22}  
\frac{\mu^2}{T^2} 
=
1.41 \alpha^2 \frac{\hbar}{e^2}\frac{\mu^2}{T^2}
.
\end{equation}
This result is represented in Fig.~\ref{r0} by the dashed red line.

\subsection{Low temperature limit}
\label{lt1}

In the opposite limit
\[
\mu \gg T
\]
we notice that the function $I_2$ is given by Eq.~(\ref{i2-3}) with
interchanged $W$ and $x$ and can be written as
\begin{equation}
\label{i2-4}
I(x\gg 1)\approx \frac{4W}{Q} \frac{\partial}{\partial z}
\frac{\sinh x}
{\cosh zQ + \cosh x}.
\end{equation}
Then the integral over $z$ in Eq.~(\ref{gg2}) is dominated by $z\gg 1$
so that the algebraic function of $z$ in the integrand may be
approximated by unity. The corresponding contribution (\ref{gg2}) to
the non-linear susceptibility may now be approximated by the
expression \cite{fn4}
\begin{equation}
\label{ggr}
g(x, |W|<Q) = 4
\frac{W}{Q} \sqrt{1-\frac{W^2}{Q^2}}
\frac{\sinh x}
{\cosh Q + \cosh x}.
\end{equation}
Now the momentum integral in Eq.~(\ref{f0-2}) is logarithmic and is
dominated by large values of momentum $Q\gg W$. Then the function
(\ref{f0-2}) takes the form
\begin{equation}
\label{f0_lt}
f_0(x,x) = \frac{64}{\pi^2} \int\limits_0^\infty \frac{dW W^2}{\sinh^2W}
\int\limits_W^\infty \frac{dQ}{Q}
\frac{\sinh^2x}{\left(\cosh Q + \cosh x\right)^2}.
\end{equation}
The ratio of the hyperbolic functions in Eq.~(\ref{f0_lt}) is similar
to the step function: it's equal to unity for $Q\ll x$ and vanishes at
larger values of momentum $Q\gg x$. Therefore $x$ effectively acts as
the upper cut-off and the momentum integral can be approximated by a
logarithm
\[
\int\limits_W^\infty \frac{dQ}{Q}
\frac{\sinh^2x}{\left(\cosh Q + \cosh x\right)^2}
\approx \ln \frac{x}{W}.
\]
Therefore, in the low temperature limit the leading contribution to
the drag conductivity is logarithmic in the chemical potential and
quadratic in temperature
\begin{equation}
\label{sd_lt}
\sigma_D(T\ll\mu) \approx \frac{32}{3} \; \alpha^2 e^2 T^2\tau^2 
\ln \frac{\mu}{T}.
\end{equation}

The single-layer conductivity at low temperatures is determined by the
chemical potential \cite{po2} and thus is given by the Drude formula
\begin{equation}
\label{s0lt}
\sigma_0 \approx (2/\pi) \; e^2\mu\tau.
\end{equation}
Consequently the drag coefficient is similar to Eq.~(\ref{d2}). 
\begin{equation}
\label{rdlt}
\rho_D (\mu\gg T)\approx  
\alpha^2\frac{\hbar}{e^2} 
\frac{8\pi^2}{3} 
\frac{T^2}{\mu^2} \ln \frac{\mu}{T}.
\end{equation}
This is to be expected, since at low temperatures $T\ll\mu$ the
phase-space argument yielding the $T^2$ dependence is justified and
the electron-hole asymmetry determines the dependence on the chemical
potential. The logarithmic factor is of course beyond such qualitative
estimates.

This result (\ref{rdlt}) is represented in Fig.~\ref{r0} by the dashed
green line and is shown together with the result of the direct
numerical calculation (shown by the solid blue line). Note that the
drag conductivity (\ref{sd_lt}) was calculated with logarithmic
accuracy.

\section{Asymptotic behavior of the drag between inequivalent layers}
\label{ar2}

Consider now the more realistic \cite{tut} situation where the two
layers are characterized by different chemical potentials. We will
still assume that the temperatures of the two layers are the same.

Note that in this Section the subscripts of the chemical potentials
$\mu_1$ and $\mu_2$ do not indicate the passive and active layers, but
rather simply distinguish between two layers with different carrier
density.

\subsection{One layer near the Dirac point}

Firstly, suppose that one of the two layers is characterized by a
small chemical potential
\[
\mu_1 \ll T.
\]
Then in Eq.~(\ref{f0-2}) the function $g(x_1)$ may be expanded using
Eq.~(\ref{i2-3}):
\[
\sigma_D = \alpha^2 e^2 \mu_1 T \tau^2 f_1\left(\frac{\mu_2}{T}\right).
\]
The dimensionless function $f_1(x)$ is characterized by the 
two limits. If the second layer is also close to the Dirac point, then
\[
f_1(x\ll 1) \approx {\cal N}_1 x,
\]
which yields a straightforward generalization of Eq.~(\ref{f0_dp}):
$f_0(x_1, x_2)\approx {\cal N}_1 x_1 x_2$. If, on the other hand the
chemical potential of the second layer is large compared to $T$, then
the drag conductivity (\ref{f0-2}) can be calculated by using the
linear approximation (\ref{i2-3}) for $g(x_1)$ and the low temperature
approximation (\ref{ggr}) for $g(x_2)$. Due to the exponential decay
of Eq.~(\ref{i2-3}) at $Q\gg W$, the momentum integral is dominated by
$Q\sim 1$ and thus the combination of the hyperbolic functions in
Eq.~(\ref{ggr}) can be replaced by unity. Thus in this limit the drag
conductivity in independent of the chemical potential of the second
layer and
\begin{equation}
\label{f1g}
f_1(x\gg 1) \approx {\cal N}_2,
\quad
{\cal N}_2 \approx 3.26.
\end{equation}

Now, the single-layer conductivity in the first layer is given by
Eq.~(\ref{s0dp}), while in the second layer we should consider both
limits. If $\mu_2$ is also small then the resulting drag
coefficient is a trivial generalization of Eq.~(\ref{rddp}):
\begin{equation}
\label{rd-12s}
\rho_D(\mu_1,\mu_2\ll T) \approx 1.41 \alpha^2 \frac{\hbar}{e^2} 
\frac{\mu_1\mu_2}{T^2}.
\end{equation}
In the opposite limit the drag conductivity is independent of the
properties of the second layer, since the integrals in
Eq.~(\ref{f0-2}) are dominated by the region where both frequency and
momentum are of order $T$. The single-layer conductivity in the second
layer however is determined by $\mu_2$ [see Eq.~(\ref{s0lt})] and thus
\begin{equation}
\label{rd-12}
\rho_D(\mu_1\ll T\ll \mu_2) \approx \alpha^2
 \frac{\pi^2{\cal N}_2}{8\ln 2} \frac{\hbar}{e^2}
\frac{\mu_1}{\mu_2}
= 5.8\alpha^2\frac{\hbar}{e^2}
\frac{\mu_1}{\mu_2}.
\end{equation}

\subsection{One layer with high carrier density}

Secondly, if the chemical potential in one of the layers is much
larger than temperature (without loss of generality we may also assume
that the chemical potential of the other layer is also smaller)
\[
\mu_2 \gg T, \quad \mu_2 > \mu_1,
\]
then arguing along the lines of the previous subsection we find that
the drag conductivity is independent of the largest chemical potential
\[
\sigma_D = \alpha^2 e^2 T^2 \tau^2 f_2(\mu_1/T).
\]
Again, we characterize the dimensionless function $f_2(x)$ by the two
limits. The situation when the second layer is near its Dirac point
was already discussed in the previous subsection. Therefore, similar
to Eq.~(\ref{f1g})
\[
f_2(x\ll 1) \approx {\cal N}_2 x.
\]
The drag coefficient in this case is given by
Eq.~(\ref{rd-12}). 
%with the obvious change of notations.

In the case where $\mu_2>\mu_1\gg T$ the calculation is similar to
that presented in Sec.~\ref{lt1}. However now the combination of the
hyperbolic functions [coming from the approximation (\ref{ggr})] in
the momentum integral similar to Eq.~(\ref{f0_lt}) comprises two
step functions and the integration is cut off by the smaller 
chemical potential. Hence
\[
f_2(x\gg 1) \approx \frac{32}{3}\ln \frac{\mu_1}{T},
\]
the drag conductivity is independent of the larger chemical
potential $\mu_2$. The drag coefficient in the limit of large $\mu_2$
(but smaller than $\mu_1$) is a generalization of
Eq.~(\ref{rdlt})
\begin{equation}
\label{rdlt-2}
\rho_D (\mu_2>\mu_1\gg T)\approx \alpha^2 \frac{\hbar}{e^2}
\frac{8\pi^2}{3} 
\frac{T^2}{\mu_1\mu_2} \ln \frac{\mu_1}{T}.
\end{equation}

\section{Beyond the lowest order perturbation theory}
\label{al}

\subsection{Finite inter-layer spacing and static screening}
\label{fd}

Let us now discuss what happens for non-zero inter-layer spacing
$d$. For simplicity, we will consider the case of two identical
layers. Generalization to the case of inequivalent layers is achieved
along the lines of Sec.~\ref{ar2}.

\subsubsection{Vanishing interaction strength}
\label{fd1}

In the limit $\alpha\rightarrow 0$ the inter-layer spacing appears
only in the exponential factor in the unscreened inter-layer
interaction (\ref{bc}). Therefore, the momentum integration acquires a
firm upper cut-off $q<v/d$, but the behavior at small momenta is
unchanged.

As we have seen in Sec.~\ref{ar}, the behavior of the drag coefficient
in the vicinity of the Dirac point is determined by frequencies and
momenta of the order of $T$ (within our perturbation theory the
single-layer conductivities in the two layers are independent of each
other and $d$). Under the assumption (\ref{td}) these momenta are
small compared to $v/d$: taking into account the exponential in
Eq.~(\ref{bc}) yields a small correction to the numerical coefficient
in Eqs.~(\ref{rddp}) and (\ref{rd-12s}): $1.41 \rightarrow 1.41 +
  4.06 Td/v$.

In the opposite limit of low temperature we have found the behavior of
the drag conductivity that is determined by large momenta. In the
result (\ref{rdlt}) the upper limit of the logarithm is given by the
chemical potential [due to the step-like behavior of the non-linear
susceptibility (\ref{ggr})], while the lower limit is given by
temperature as the typical value of frequency in Eq.~(\ref{f0_lt}).

Clearly, if finite $d$ is taken into account then the logarithmic
behavior in Eq.~(\ref{rdlt}) may change: the upper limit of the
logarithm will now depend on the relative value of the chemical
potential and inverse inter-layer spacing.

At the same time the lower limit of the logarithm also depends on our
approximations. Indeed, so far we have considered the effect of the
unscreened inter-layer interaction. Consider now the role of the
static screening. If both $d$ and $\alpha$ are small then we may
approximate the inter-layer interaction (\ref{d12}) by the expression
\begin{equation}
\label{d12p}
{\cal D}^R_{12} = - 
%\frac{2\pi e^2}{q}
%\frac{e^{-qd}}{1+4\pi e^2 \Pi^R/q}.
\frac{e^{-qd}}{\displaystyle\frac{q}{2\pi e^2} + 2\Pi^R}.
%\frac{e^{-qd}}{q/(2\pi e^2) + 2\Pi^R}.
\end{equation}
At low temperatures the leading contribution to the static
polarization operator is
\begin{equation}
\label{po_lt}
\Pi^R(\omega=0) = \frac{2k_F}{\pi v}.
\end{equation}
Thus the interaction propagator can be written as
\[
{\cal D}^R_{12} = - \frac{2\pi\alpha v^2}
{vq+2N\alpha\mu } \;
e^{-qd}.
\]
Here $N=4$ is due to spin and valley degeneracy. The unscreened
interaction is a good approximation as long as the inverse screening
length is small compared to typical momenta, i.e. as long as
$N\alpha\mu\ll T$. In the opposite regime the lower limit of the
logarithm in Eq.~(\ref{sd_lt}) will be determined by the screening
length.

{
\begin{figure}
\begin{center}
\epsfig{file=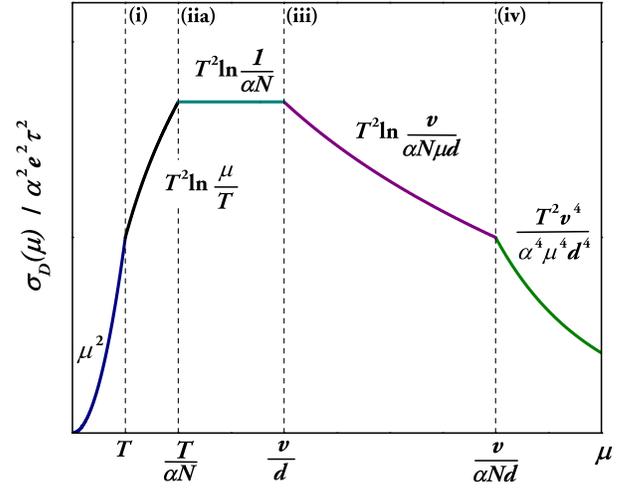,width=8cm}
\end{center}
\caption{[Color online] The sketch of the drag conductivity (in the
  units of $\alpha^2 e^2 \tau^2$) as a function of the chemical
  potential illustrating the results of Section~\ref{fd1} in the case
  $T\ll N\alpha v/d$. The blue line corresponds to the quadratic
  dependence (\ref{sd-pt-l1}) in the vicinity of the Dirac point. The
  four regions at large chemical potentials correspond to the four
  regions discussed in Section~\ref{fd1}. If $T\gg N\alpha v/d$, then
  the region (iia) should be replaced by (ii), the logarithmic
  dependence $T^2\ln(1/\alpha)$ should be replaced by $T^2\ln[v/(Td)]$, 
  and the limits $v/d$ and $T/(N \alpha)$ should be exchanged.}
\label{sk1}
\end{figure}
}

Combining the above arguments we conclude that increasing the chemical 
potential the following four regimes may be gradually achieved.

(i) $N\alpha\mu\ll T\ll\mu\ll v/d$. This regime was considered in
Sec.~\ref{ar} leading to Eqs.~(\ref{sd_lt}) and (\ref{rdlt}).

(ii) $N\alpha\mu\ll T\ll v/d\ll\mu$. If the chemical potential is
increased beyond the inverse inter-layer spacing, then the momentum
integration in Eq.~(\ref{f0_lt}) is cut off by $v/d$ instead of $\mu$.
The logarithmic behavior of the drag conductivity Eq.~(\ref{sd_lt})
will be modified and $\sigma_D$ no longer depends on the chemical 
potential
\begin{equation}
\label{s2}
\sigma_D \sim \alpha^2e^2T^2\tau^2\ln\frac{v}{Td}.
\end{equation}

(iia) $T\ll N\alpha\mu\ll\mu\ll v/d$. Depending on the actual values 
of $T$, $d$, and $\alpha$ it is possible that $N\alpha\mu$ exceeds
temperature while the chemical potential is still smaller than the
inverse inter-layer spacing $\mu\ll v/d$. Then instead of the previous
regime we find 
\begin{equation}
\label{s2a}
\sigma_D \sim \alpha^2e^2T^2\tau^2\ln\frac{1}{N\alpha}.
\end{equation}

(iii) $T\ll N\alpha\mu\ll v/d\ll\mu$. Increasing the chemical
potential further leads to the regime where the static screening can
no longer be neglected. Now the lower integration limit in
Eq.~(\ref{f0_lt}) is effectively given by the inverse screening length
rather than the frequency. The upper limit is still determined by the
inter-layer spacing. Therefore the drag conductivity again depends
logarithmically on the chemical potential \cite{kac}
\begin{equation}
\label{s3}
\sigma_D \sim \alpha^2e^2T^2\tau^2\ln\frac{v}{N\alpha\mu d},
\end{equation}
but now this is a {\it decreasing} function, indicating the existence
of the absolute maximum of the drag conductivity as a function of the
chemical potential.

(iv) $T\ll v/d \ll N\alpha\mu \ll\mu$. Finally, if the chemical
potential is so large that the screening length becomes smaller than
the inter-layer spacing the momentum integral in Eq.~(\ref{f0_lt}) is
no longer logarithmic. In this regime we recover the standard
Fermi-liquid result\cite{kor}. Note, that in this regime the step-like
combination of the hyperbolic function in the non-linear
susceptibility (\ref{ggr}) is completely ineffective and may be
replaced by unity. Given that in this regime the integration is
dominated by momenta large compared to temperature the non-linear
susceptibility may be further linearized in frequency. The resulting
expression 
\begin{equation}
\label{ft}
\sigma_D = \frac{\zeta(3)}{4}
\frac{e^2\tau^2T^2}{(k_Fd)^2(\varkappa d)^2}, \quad \varkappa = 4\alpha k_F,
\end{equation}
differs from that of Ref.~\onlinecite{kor} only by the factor
reflecting valley degeneracy in graphene. Such expression for the drag
conductivity was previously obtained in Ref.~\onlinecite{dsa}. The
results of this subsection are illustrated in Fig~\ref{sk1}.

\subsubsection{Intermediate interaction strength and $\mu\gg T$}
\label{fd2}

The above results rely on the smallness of the interaction strength
$\alpha$. However, if $N\alpha>1$, then (i) the
approximation (\ref{d12p}) might not be justified and we would need to
use the full expression (\ref{d12}) for the interaction propagator;
(ii) the four regimes specified in the previous subsection may not
exist, since it might happen that $T/(N\alpha)\ll T<v/(N\alpha d)\ll
v/d$. In this case there are only two distinctive regimes (for $\mu\gg
T$): (a) $\mu\ll v/d$, and (b) $\mu\gg v/d$. The latter regime is
usually identified with the Fermi-liquid result
\cite{kor,tut,dsa,kac,csn}, i.e. Eq.~(\ref{ft}).

In this subsection we derive the approximate expression for the drag
conductivity for large values of the chemical potential $\mu\gg T$,
which accounts for the possibility of $N\alpha>1$ (possible for small
$\alpha\ll 1$, but large $N\gg 1$; here we still consider identical
layers). In this regime the single-layer conductivity (\ref{s0lt}) is
large (since $\mu\tau\gg T\tau \gg 1$) and therefore we can still limit
our consideration to the diagram in Fig.~\ref{diagram}. Moreover, 
for $\mu\gg T$ we can somewhat relax the condition (\ref{att}) on the
interaction strength and require the electron-electron scattering time
to be larger than the impurity scattering time
\[
\tau_{ee}\gg\tau \quad \Rightarrow \quad
\tau^{-1} \gg \; \alpha^2 \frac{T^2}{\mu}
\quad \Rightarrow \quad
\alpha^2T\tau \ll \frac{\mu}{T}.
\]
Now we can follow the usual steps leading to the Fermi-liquid result
(\ref{ft}). We consider only the static screening by approximating the
polarization operator by Eq.~(\ref{po_lt}) (see Appendix~\ref{pol_op}
for details). We further assume \cite{fn4} that the dominant
contribution to the drag conductivity comes from the region
$vq>\omega$. Finally, we assume that in that region the result of the
momentum integral is determined by the upper integration limit and is
therefore independent of $\omega$. This allows us to evaluate the
frequency integral and represent the drag conductivity in terms of the
single integral over momenta. Under these assumptions we find
similarly to Eqs.~(\ref{f0_lt})
\begin{subequations}
\label{r1}
\begin{equation}
\sigma_D = \alpha^2 e^2 T^2 \tau^2 
f_0\left(\frac{\mu}{T}; \alpha; \frac{Td}{v}\right),
\end{equation}
\begin{eqnarray*}
&&
f_0(x; \alpha ;\lambda) \approx \frac{32}{3}
\int\limits_{1}^\infty 
\frac{dQ Q^3 e^{-4\lambda Q}}
{\left[(Q+\tilde\alpha(x))^2 - \tilde\alpha(x)^2e^{-4\lambda Q}\right]^2}
\nonumber\\
&&
\nonumber\\
&&
\quad\quad\quad\quad\quad
\quad\quad\quad\quad\quad
\times
\frac{\sinh^2x}{\left(\cosh Q + \cosh x\right)^2},
\end{eqnarray*}
where
\begin{equation}
\label{at}
\tilde\alpha(x) = 
\frac{N}{2} \; 
%N
\alpha x
%/2
.
\end{equation}
Here we have approximated the function $g(x, |W|<Q)$ by
Eq.~(\ref{ggr}), but neglected the frequency under the square root. In
the limit $\mu\gg T$ the combination of the hyperbolic functions in
$g(x, |W|<Q)$ has the form of the step function, which effectively
cuts off the integration at $Q\sim x$.  Thus we arrive at the
approximate expression
\begin{equation}
f_0(x; \alpha ;\lambda) \approx \frac{32}{3}
\int\limits_{1}^x
\frac{dQ Q^3 e^{-4\lambda Q}}
{\left[(Q+\tilde\alpha(x))^2 - \tilde\alpha(x)^2e^{-4\lambda Q}\right]^2}.
\end{equation}

The results of the previous subsection can now be recovered for
$\alpha\ll 1$ by neglecting the terms proportional to $\tilde\alpha^2$
in the denominator. In contrast, the Fermi-liquid result (\ref{ft})
can be obtained by (i) assuming $\tilde\alpha\gg 1$ and keeping only the
terms proportional to $\tilde\alpha^2$ in the denominator, (ii) assuming
$x\gg 1/4\lambda$ and thus replacing the upper integration limit by
infinity, and (iii) replacing the lower integration limit by zero.
Moreover, for $\mu\gg v/d$ the integration limit can be extended such that
the result becomes a function of one single parameter
\begin{equation}
f_0(x; \alpha ;\lambda) \approx \tilde f_0(4\lambda\tilde\alpha),
\end{equation}
where
\begin{equation}
\label{rf0}
\tilde f_0(y) = \frac{32}{3}
\int\limits_{0}^\infty
\frac{dZ Z^3 e^{-Z}}
{\left[(Z+y)^2 - y^2e^{-Z}\right]^2}.
\end{equation}
\end{subequations}
The function (\ref{rf0}) describes the crossover between the regimes
(iii) and (iv) of the previous subsection (see Fig.~\ref{sk1}). 
This can be seen by evaluating the integral in the two limits
(here $\gamma_0\approx 0.577216$ is the Euler's constant)
\begin{subequations}
\begin{equation}
\tilde f_0(y\ll 1) \approx \frac{32}{3} \left(\ln\frac{1}{y}
-\gamma_0-\frac{11}{6}\right),
\end{equation}
\begin{equation}
\tilde f_0(y\gg 1) \approx \frac{64\zeta(3)}{y^4}.
\end{equation}
\end{subequations}

It turns out that by numerical reasons this crossover spans a large
interval of values of the chemical potential such that the
Fermi-liquid result (\ref{ft}) is practically unattainable in
graphene-based drag measurements \cite{tut}. We illustrate this point
in Figs.~\ref{cp1}~-~\ref{cp3}.

{
\begin{figure}
\begin{center}
\epsfig{file=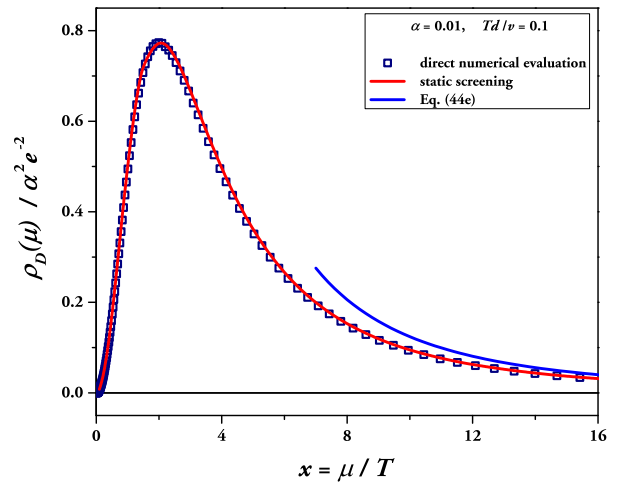,width=8cm}
\end{center}
\caption{[Color online] Results of the numerical evaluation of the
  drag coefficient for $\alpha=0.01$ and $Td/v=0.1$. The squares
  represent the calculation of Eq.~(\ref{sd}) with the only
  approximation that the polarization operator in the screened
  inter-layer interaction (\ref{d12}) was evaluated in the absence of
  disorder. The red line corresponds to the same calculation, but
  replacing the polarization operator by Eq.~(\ref{po_lt}),
  i.e. taking into account only static screening. The blue line
  corresponds to the approximate expression (\ref{rf0}),valid for
  $\mu\gg v/d$.}
\label{cp1}
\end{figure}
}

{
\begin{figure}
\begin{center}
\epsfig{file=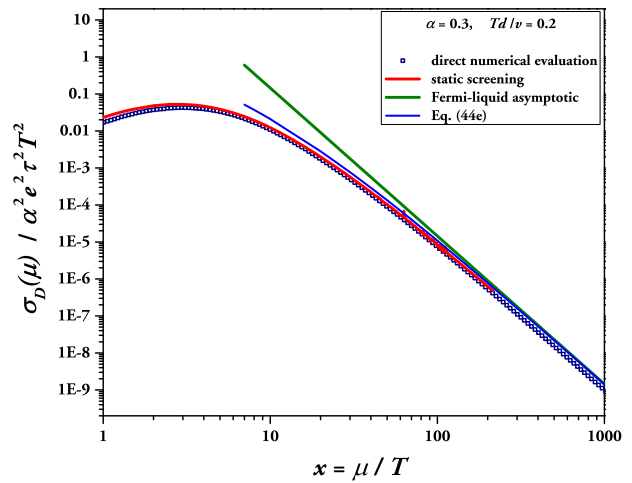,width=8.2cm}
\end{center}
\caption{[Color online] Results of the numerical evaluation of the
  drag conductivity for $\alpha=0.3$ and $Td/v=0.2$ shown in the
  log-log scale. The straight green line represents the Fermi-liquid
  result (\ref{ft}). The solid blue line corresponds to the
  approximate expression (\ref{rf0}),valid for $\mu\gg v/d$.}
\label{cp3}
\end{figure}
}

First of all we evaluate the whole expression (\ref{sd}). In order
to do that we need to evaluate the full non-linear susceptibility
(\ref{gg}). In addition we need to determine the polarization operator
in graphene at finite temperature and chemical potential. In the
ballistic regime we neglect the effect of disorder on the polarization
operator. This is the only approximation in this calculation, see
Appendix~\ref{pol_op} for details. The role of the disorder is
discussed in the following subsections. The corresponding data is
shown in Figs.~\ref{cp1}~-~\ref{cp3} by blue squares. Further results
of the numerical evaluation of the drag conductivity (\ref{sd}) are
shown in Appendix~\ref{numres}.

Then we can simplify calculations by using the static approximation
(\ref{po_lt}) for the polarization operator. The result of this
calculation is shown in Figs.~\ref{cp1}~-~\ref{cp3} by the red line.
For weak interaction $\alpha=0.01$ and relatively small inter-layer
spacing (such that $Td/v=0.1$) the results of the static screening
approximation are indistinguishable from the ``full'' calculation as
can be seen in Fig.~\ref{cp1}. At the same time, for ``intermediate''
values\cite{tut,csn} $\alpha=0.3$, $Td/v=0.2$, the static screening 
approximation ``works'' well for small and large chemical potentials, 
while somewhat overestimating the overall peak height.

Finally, we can evaluate the approximate expression (\ref{rf0}).  This
expression was derived assuming $\mu\gg v/d$. For the parameter values
used in Figs.~\ref{cp1}~-~\ref{cp3} the expression (\ref{rf0}) fits
the exact calculation for $\mu/T>10$. However, the Fermi-liquid
asymptotic (represented by the straight green line in Fig.~\ref{cp3})
is not reached until $\mu/T>200$. Clearly this is very far from the
parameter regime relevant to the experimental observation of the
Coulomb drag in graphene \cite{tut,csn}. It is therefore not
surprising that the Fermi-liquid-like approximations overestimate the
observed values of the drag \cite{tut}. The data in Fig.~\ref{cp3}
show that the asymptotic results, such as Eq.~(\ref{ft}) may sometimes
be achieved only at the extreme values of parameters. In order to
describe the effect in the intermediate (or realistic) parameter
regime, one needs to evaluate Eq.~(\ref{sd}) with only minimal
approximations.

\subsection{Energy-dependent scattering time}
\label{mft}

The drag coefficient (\ref{r0-1}) and the asymptotic results of
Sec.~\ref{ar} and Sec.~\ref{ar2}, as well as the numerical data of
Sec.~\ref{fd}, were obtained assuming a constant impurity scattering
time $\tau$. In graphene, the scattering time strongly depends on the
type of the impurities and usually depends on energy
\cite{tac,tau,tan,tas,ta2}. In the context of the Coulomb drag in
graphene a similar issue (namely, the momentum-dependent scattering
time) was investigated in Ref.~\onlinecite{csn} in the framework of
the kinetic equation.

\subsubsection{Non-linear susceptibility}

Consider now the effect of the energy (or momentum) dependence of the
scattering time. Within our assumption (\ref{att}) we can still
consider the ballistic Green's functions (\ref{gdel}), only now
instead of being the overall factor, the scattering time is a part of
the integrand in the non-linear susceptibility (\ref{nls}). Due to the
$\delta$-function form of the Green's functions we can focus on the
energy-dependent $\tau$ regardless of the microscopic impurity model.

Repeating the steps leading to Eq.~(\ref{gg}) with the
energy-dependent $\tau(\epsilon)$ we find that the functions
(\ref{gg1}) and (\ref{gg2}) become (see Appendices~\ref{gamma} and
\ref{kin_ur} for details)
\begin{subequations}
\label{gt}
\begin{equation}
g(x) = \left\{
\begin{matrix}
\displaystyle\int\limits_0^1 dz 
\displaystyle\frac{\sqrt{1-z^2}}{2\sqrt{\frac{W^2}{Q^2}-1}} \; 
I(z) K(z), & |W|>Q, \cr\cr
\displaystyle\int\limits_1^\infty dz 
\displaystyle\frac{\sqrt{z^2-1}}{2\sqrt{1-\frac{W^2}{Q^2}}}\;
I(z) K(z), & |W|<Q,
\end{matrix}
\right.
\label{gtk}
\end{equation}
where
\begin{eqnarray}
\label{k1}
&&
K(z) = \frac{zW-Q}{zQ-W}\frac{\tau(T[zQ-W])}{\tau(T)} 
\\
&&
\nonumber\\
&&
\quad\quad\quad\quad\quad\quad\quad\quad
- 
\frac{zW+Q}{zQ+W}\frac{\tau(-T[zQ+W])}{\tau(T)},
\nonumber
\end{eqnarray}
\end{subequations}
so that the explicit factor of the scattering time in Eq.~(\ref{gg0})
should be understood as $\tau(T)$. This choice of the prefactor is not
essential and is dictated by the discussion of the case $\mu\ll T$
below.

In the simple case $\tau(\epsilon)=const$, the function $K(z)$
simplifies to
\begin{equation}
\label{k}
K(z) = - 2z \frac{1-W^2/Q^2}{z^2-W^2/Q^2},
\end{equation}
and we recover Eqs.~(\ref{gg1}) and (\ref{gg2}).

\subsubsection{Vicinity of the Dirac point}

Taking into account energy dependence of the scattering time
$\tau(\epsilon)$ modifies the non-linear susceptibility and thus
changes the Coulomb drag. In the limit $\mu\ll T$ the region
$Q\sim W$ becomes important. As can be seen from Eqs.~(\ref{gg1}) and
(\ref{gg2}), if $\tau(\epsilon)=const$, then precisely at $Q=|W|$ the
non-linear susceptibility vanishes. Otherwise, Eqs.~(\ref{gt}) may
contain a divergence.

Consider for example Coulomb scatterers. Then \cite{tac,tau}
\begin{equation}
\tau(\epsilon) = \tau_0^2 |\epsilon|.
\end{equation}
In this case the function (\ref{k}) depends on the relation between
$|W|$ and $Q$
\begin{equation}
K(z) \rightarrow - 2
%T\tau_0^2 
\left\{
\begin{matrix}
Q,  &  Q>W, \cr
z |W|, & Q<W,
\end{matrix}
\right.
\label{kc}
\end{equation}
and as a result the integral (\ref{f0-2}) contains a logarithmic
divergence at $Q=|W|$ since now
\begin{equation}
\label{div}
%g(x_1) g(x_2)\propto\frac{1}{|W^2-Q^2|}.
g(x_1) g(x_2)\propto|W^2-Q^2|^{-1}.
\end{equation}
Same result holds for the case of strong short-ranged impurities that
also yield the linear dependence $\tau\sim|\epsilon|$ (up to
logarithmic renormalization which is inessential in the present
context) \cite{ta2}.

Similarly, in the case of weak short-ranged disorder \cite{tan}
\begin{equation}
\tau^{-1}(\epsilon) = \gamma |\epsilon|,
\end{equation}
the function (\ref{k1}) does not vanish \cite{fn3} for $Q=|W|$:
\[
K(z) \rightarrow - 2
%\frac{2}{\gamma T}
\left\{
\begin{matrix}
Q\displaystyle\frac{z^2Q^2 + W^2 - 2z^2W^2}{(z^2Q^2 - W^2)^2}, &  Q>|W|, \cr\cr
z|W|\displaystyle\frac{z^2Q^2 + W^2 - 2Q^2}{(z^2Q^2 - W^2)^2}, & Q<|W|.
\end{matrix}
\right.
\]
Therefore the non-linear susceptibility contains the same logarithmic
divergence (\ref{div}). 

Finally, even in the case of logarithmic renormalization of the
scattering time \cite{tas,ta2} the function (\ref{k}) does not vanish
for $Q=|W|$ leading to Eq.~(\ref{div}). We conclude, that
Eq.~(\ref{div}) is the generic behavior, while the vanishing (for
$Q=|W|$) non-linear susceptibility (\ref{gg}) is the artifact of the
approximation $\tau(\epsilon)=const$.

The divergence indicates that in the region $W\sim Q$ the
$\delta$-function approximation for the Green's functions (\ref{gdel})
is invalid. Going beyond this approximation is qualitatively
equivalent to regularizing the divergence by the scattering time
$\tau(T)$. Thus the drag coefficient in the vicinity of the Dirac
point acquires an additional logarithmic factor
\begin{equation}
\label{rd-12t}
\rho_D(\mu_1,\mu_2\ll T) \sim \alpha^2 \frac{\hbar}{e^2} 
\frac{\mu_1\mu_2}{T^2} \ln T\tau(T).
\end{equation}

\subsubsection{Low temperature limit}

In the low temperature limit the non-linear susceptibility is
determined by electrons near the Fermi surface and one can set
$\epsilon$ in the triangular vertex (\ref{g}) to be equal to the
chemical potential. Then it appears that in all subsequent expressions
one has to replace $\tau(\epsilon)$ by $\tau(\mu)$
\begin{equation}
\label{taum}
\mu\gg T \quad\Rightarrow\quad
\tau(\epsilon)\approx\tau(\mu).
\end{equation}

Let us understand this statement in more detail. Consider for example
Coulomb scatterers. In the limit $\mu\gg T$ the non-linear
susceptibility is dominated by $Q\gg W$. In this case we can evaluate
the integral in Eq.~(\ref{gtk}) using the function $K(z)$ from
Eq.~(\ref{kc}) and approximating $I(z)$ by Eq.~(\ref{i2-4}). As a
result we find
\begin{eqnarray}
&&
g(Q\gg W) = 
\frac{4W\sinh x}{\cosh Q + \cosh x} +
\frac{4W}{Q} \ln\frac{e^{2x}+e^{x+Q}}{1+e^{x+Q}}
\nonumber\\
&&
\nonumber\\
&&
\quad\quad\quad\quad\quad\quad
\simeq 4W \left[1 + \frac{x}{Q}\right]
\theta(x-Q).
\label{gc}
\end{eqnarray}
The momentum integral is still logarithmic. Therefore we can approximate
$g(Q\gg W)$ by
\[
g(Q\gg W) \approx \frac{4W}{Q} x,
\]
which indeed recovers the statement (\ref{taum}) 
[notice that $\tau(T)x=\tau(\mu)$].

This argument is not restricted to Coulomb scatterers since in the
limit $x\gg Q$ the function $I(z)$ in Eq.~(\ref{i2-4}) is essentially
a $\delta$-function: $I(z)\approx -4W\delta(x-zQ)$. Therefore the
function $K(z)$ in the integrand in Eq.~(\ref{gt}) becomes
$K(z)\rightarrow -2\tau(\mu)/[z\tau(T)]$ for an arbitrary
$\tau(\epsilon)$. 

In addition, the function $g$ contains the divergence (\ref{div}) at
$Q\sim W$ which is always present for $\tau(\epsilon)\ne
const$. Similarly to Eq.~(\ref{rd-12t}) this results in an additional
logarithmic contribution (\ref{div}), which is typically
subleading. For example, the Fermi-liquid result (\ref{ft}) acquires
an additional factor $1 + (Td/v)\ln T\tau(T)$. Thus we conclude that
for $\mu\gg T$ taking into account the energy dependence of the
impurity scattering time does not affect our results.

Finally, the divergence (\ref{div}) appears only if both chemical
potentials are small. Indeed, if $\mu_1\ll T\ll\mu_2$, then while the
non-linear susceptibility in the first layer is given by
Eq.~(\ref{gt}), in the second layer the scattering time may be
replaced by its value at the chemical potential and thus
Eq.~(\ref{gg}) applies. Now the divergent denominator
$|W^2-Q^2|^{-1/2}$ of the former susceptibility is canceled by
precisely the same factor in the numerator in the latter. Hence, the
product $g(x_1)g(x_2)$ is finite at $|W|=Q$.

\subsection{Plasmon contribution}

The above results were obtained while neglecting the plasmon pole of
the interaction propagator. Let's now estimate the plasmon
contribution keeping $\alpha$ small. Consider for simplicity identical
layers. In the limit $d\rightarrow 0$, the plasmon pole in the
propagator of the inter-layer interaction is very similar to that of
the single layer. Indeed, introducing dimensionless functions
\begin{equation}
\label{dn}
{\cal D}^R_{12} = - \frac{2\pi e^2}{q} D, \quad
\Pi^R = \frac{2q}{\pi^2 v} P,
\end{equation}
the inter-layer propagator (\ref{d12}) can be expressed as
\[
D^{-1}=(1+\beta P)^2e^{qd} - \beta^2 P^2 e^{-qd}, \quad 
\beta = \frac{4\alpha}{\pi}.
\]
In the limit $d\rightarrow 0$ the pole corresponds to the solution of
the equation
\[
1+2\beta P =0,
\]
which differs from its single-layer counterpart by the factor of $2$
only.

Now the drag conductivity (\ref{sd}) is determined by the product of
the square of the interaction propagator and two non-linear
susceptibilities. Using the dimensionless notations (\ref{dn}) and
(\ref{gg}) we can write the integrand in Eq.~(\ref{sd}) in the form
\begin{equation}
\label{ggdd}
\Gamma^\alpha_1\Gamma^\beta_2 |{\cal D}^R_{12}|^2 \propto
e^2\tau^2 \alpha^2 \frac{q^\alpha q^\beta}{q^2} \; g^2 |D|^2.
\end{equation}
Calculations of Sec.~\ref{cdpt} are essentially equivalent to
arguments based on the Fermi Golden Rule yielding the perturbative
result $\rho_D\sim\alpha^2$. Now we show that taking into account 
the plasmon contribution results in additional smallness, justifying the
perturbative calculation of Sec.~\ref{cdpt}.

Plasmon modes in a single graphene sheet in the vicinity of the Dirac
point were studied in Ref.~\onlinecite{po2}. The plasmon pole appears
in the region 
\[
W>Q.
\]
Adjusting for the above factor of $2$, the plasmon dispersion and
decay rate are given by
\begin{equation}
\label{pdis}
W_p = \sqrt{\frac{Q}{Q+2\tilde\alpha}} \; (Q+\tilde\alpha), \quad
\tilde\alpha = 8\alpha\ln 2,
\end{equation}
\begin{equation}
\Gamma_p = \frac{\pi \tilde\alpha^2}{16 \ln 2} 
\left(\frac{Q}{Q+2\tilde\alpha}\right)^{3/2}(Q+\tilde\alpha).
\end{equation}

The plasmon contribution to the drag conductivity may be described by
the following form of the inter-layer interaction propagator
\[
|D|^2 \sim \frac{(W^2-Q^2)^3}{\tilde\alpha^2Q^2}
\frac{1}{(W-W_p)^2+\Gamma_p^2}.
\]
At the same time, as argued in Ref.~\onlinecite{po2}, the typical 
momenta dominating the relaxation rates are not too small
\begin{equation}
\label{tm}
Q \gtrsim \tilde\alpha.
\end{equation}
Then the decay rate is small $\Gamma_p\sim\tilde\alpha^3$ and the
interaction propagator has the form of a sharp peak. Estimating the 
typical frequency at the peak by the plasmon dispersion we note that
\begin{equation}
\label{es1}
W_p^2-Q^2 = \frac{\tilde\alpha^2 Q}{Q+2\tilde\alpha} \sim 
\tilde\alpha^2,
\end{equation}
and therefore the inter-layer interaction takes the form (omitting
inessential numerical factors)
\[
|D|^2 \sim 
\frac{\tilde\alpha^2}{(W-W_p)^2+\tilde\alpha^6}.
\]

Consider now the non-linear susceptibility $g$.
Typical momenta (\ref{tm}) are small enough, which allows us to
approximate the non-linear susceptibility (\ref{gg1}) by its
asymptotic value in the small $Q\rightarrow 0$ limit
\begin{equation}
\label{ggq0}
g_1(Q\rightarrow 0) \approx \frac{\pi Q^2}{4W}.
\end{equation}
As a result, for $Q\sim\tilde\alpha$,
\[
g^2|D|^2 \propto 
\frac{\tilde\alpha^6}{(W-W_p)^2+\tilde\alpha^6}
\rightarrow \tilde\alpha^3 \delta(W-W_p).
\]

For larger momenta $Q\sim 1$ the result is similar. The interaction
propagator is now given by
\[
|D|^2 \sim 
\frac{\tilde\alpha^4}{(W-W_p)^2+\tilde\alpha^4}.
\]
At the same time the non-linear susceptibility (\ref{gg}) contains a
factor $\sqrt{W^2-Q^2}$. Using Eq.~(\ref{es1}), we find 
\[
g^2|D|^2 \propto 
\frac{\tilde\alpha^6}{(W-W_p)^2+\tilde\alpha^4}
\rightarrow \tilde\alpha^4 \delta(W-W_p).
\]

If however, the temperature dependence of the scattering time is taken
into account, then instead of vanishing at $|W|=Q$ the non-linear
susceptibility contains the divergence (\ref{div}) and thus a more
careful analysis is necessary. Now we need to find the plasmon
dispersion and decay rate at $T\ne 0$ in the presence of disorder.
This complicated problem lies beyond the scope of the present
paper. Here we estimate the plasmon contribution to $\sigma_D$ in the
most relevant region $Q\sim|W|\sim T$. It turns out that under our
assumption (\ref{att}) this contribution is small compared to the
leading approximation (\ref{r}).

In order to find the plasmon dispersion we need to calculate the
polarization operator at $T\ne 0$ in the presence of disorder. This
can be done with the help of the kinetic equation derived in
Ref.~\onlinecite{fke}. In comparison to the usual $\tau$-approximation
this equation contains an additional term describing the suppression
of backscattering in graphene. It turns out that in the region
$Q\sim|W|\sim T$ the polarization operator may be approximated by
\begin{equation}
\label{pru}
\Pi^R = \frac{\partial n}{\partial \mu} 
\left[1 - 
\frac{i\omega}{\sqrt{\left(i\omega+\frac{1}{\tau}\right)^2+v^2q^2}}
\right],
\end{equation}
where the thermodynamic density of states is given by
\begin{equation}
\label{tdos}
\frac{\partial n}{\partial \mu} = \frac{4T}{\pi v^2}
\ln\left[2\cosh \frac{\mu}{2T}\right].
\end{equation}
The absence of the extra term $1/\tau$ in the denominator (c.f.
Ref.~\onlinecite{zna}) is precisely due to the suppression of
backscattering: for arbitrary $q$ and $\omega$ there is indeed a 
rather involved expression generalizing
this term to the case of graphene, but for $Q\sim|W|\sim T$ this
contribution is small and may be neglected.

Solving for the plasmon dispersion with the help of Eq.~(\ref{pru})
we find 
\[
W_p \approx Q, \quad \Gamma_p = \tau^{-1}.
\]
Thus the plasmon dispersion is changed little from Eq.~(\ref{pdis}),
where for $Q\gg \tilde\alpha$ we also have $W_p \approx Q$. However
the decay rate is now completely determined by disorder. Now the
interaction propagator takes the form
\[
|D|^2 \sim \frac{\tilde\alpha^4 T\tau}{Q^2}
\frac{1/T\tau}{(W-W_p)^2+1/T^2\tau^2}.
\]
Multiplying this expression by the diverging $g^2$, we notice that
corrections to the linear plasmon dispersion are still determined by
the interaction as in Eq.~(\ref{es1}). Therefore, we find that the
resulting contribution contains the small factor $\alpha^2T\tau$ and
is negligible under our assumption (\ref{att}).

In order to estimate the drag conductivity we now need to integrate
the product $g^2|D|^2$ over frequency and momentum, see
Eqs.~(\ref{sd}) and (\ref{ggdd}). This product contains a small factor
of at least $\tilde\alpha^3$, or, if energy dependence of the
scattering time is taken into account, a small parameter
$\alpha^2T\tau$. Therefore the plasmons contribute in the subleading
order in the perturbative expansion in $\alpha$.

The conclusions of this Section are confirmed by comparing the results
of numerical evaluation of $\rho_D$ using either the full dynamically
screened interaction (\ref{d12}) or only the static screening, see
Eq.~(\ref{po_lt}). The results are illustrated in
Figs.~\ref{cp1}-\ref{cp3}: the difference between the two results is
only noticable for larger values of $\alpha$. Thus taking into account
plasmons does not lead to any new qualitative features of the
theory.

\subsection{Spectrum renormalization}
\label{sr}

If Coulomb interaction is taken into account, then the Dirac spectrum
in graphene acquires logarithmic corrections \cite{log}. This can be
understood in terms of the renormalization of the interaction
parameter $\alpha$ \cite{rg,rg1} and disorder strength \cite{rg2}. The
renormalization group flow terminates at ${\rm max} (\mu, T)$ and at
lower energy scales we can treat the parameters $\alpha$ and $v$ as
scale-independent and equal to their renormalized values. The disorder
scattering time retains its explicit energy dependence which follows
from the microscopic impurity model.

In our calculation of the drag conductivity all frequency integrals
are effectively cut off by temperature, while the momentum integrals
are cut off by either $T$ or $\mu$, whichever is larger. In all of
these cases we can treat the spectrum as linear with the renormalized
velocity. Then our result (\ref{rd-12t}) is still applicable, with the
velocity and interaction parameter $\alpha$ taking the renormalized
values $v[{\rm max}(\mu, T)]$ and $\alpha[{\rm max}(\mu, T)]$.

\subsection{Experimental relevance}
\label{exrel}

\subsubsection{Carrier density}

Experimental results \cite{tut} are expressed as a function of carrier
density rather than the chemical potential as we have discussed in
this paper. The relation between the carrier density $n$ and the
chemical potential $\mu$ can be obtained by integrating the density of
states $\rho(\epsilon)$:
\begin{equation}
\label{nmt}
n = \int\limits_{-\infty}^\infty d\epsilon
\rho(\epsilon) \left[n_F(\epsilon; \mu)-n_F(\epsilon; 0)\right],
\end{equation}
where $n_F(\epsilon; \mu)$ is the Fermi distribution function. 

In graphene $\rho(\epsilon)=2|\epsilon|/(\pi v^2)$ and the integral 
\begin{equation}
\label{nm}
n = \int\limits_{-\infty}^\infty \frac{d\epsilon |\epsilon|}{\pi v^2} 
\left[\tanh\frac{\epsilon}{2T} - \tanh\frac{\epsilon-\mu}{2T}\right]
\end{equation}
can be easily evaluated in the limiting cases [cf. Eq.~(\ref{tdos})]
\begin{equation}
\label{nl}
n= \frac{1}{\pi v^2}
\left\{
\begin{matrix}
\mu^2, & \mu\gg T, \cr
(4\ln 2) \mu T, & \mu\ll T, 
\end{matrix}
\right.
\end{equation}
which of course recovers the $T=0$ expression for $\mu\gg T$.

If impurity scattering is taken into account then the density of
states in the vicinity of the Dirac point saturates to a value
determined by disorder \cite{ta2} 
\begin{equation}
\label{ndis}
n(\mu, T <\tau^{-1}) = \frac{\mu}{v^2\tau}.
\end{equation}
However for $T\tau\gg 1$ this effect is not important.

In experiment the carrier density may be obtained from measurements of
the Hall coefficient in a non-quantizing magnetic field $H$. In
graphene this is more complicated than in usual metals since the Hall
coefficient vanishes at the Dirac point \cite{man} due to
electron-hole symmetry. While at low temperatures the behavior of the
Hall coefficient is rather complicated \cite{an2}, at high
temperatures we can use the conventional Boltzmann kinetic equation
with the energy-dependent cyclotron frequency \cite{an2}
$\omega_c(\epsilon) = eHv^2/(c\epsilon)$. Then we find
\begin{subequations}
\label{skin}
\begin{equation}
\label{sxx}
\sigma_{xx} = - e^2v^2 
\int d\epsilon 
\frac{\partial n_F(\epsilon)}{\partial \epsilon} 
\frac{\rho(\epsilon)\tau_{tr}(\epsilon)}
{1+\omega^2_c(\epsilon)\tau_{tr}^2(\epsilon)},
\end{equation}
\begin{equation}
\label{sxy}
\sigma_{xy} = - e^2v^2 
\int d\epsilon 
\frac{\partial n_F(\epsilon)}{\partial \epsilon} 
\frac{\rho(\epsilon)\omega_c(\epsilon)\tau_{tr}^2(\epsilon)}
{1+\omega^2_c(\epsilon)\tau_{tr}^2(\epsilon)},
\end{equation}
\end{subequations}
Using Eqs.~(\ref{skin}) at low temperatures $\mu\gg T$ and for weak
magnetic fields $H$ we of course recover the classic result
\[
R_H = - \frac{\sigma_{xy}}{(\sigma^2_{xx}+\sigma^2_{xy})H} 
= -\frac{1}{nec},
\]
with the electron density given by Eq.~(\ref{nl}).

Exactly at the Dirac point the Hall conductivity vanishes,
$\sigma_{xy}(\mu=0)=0$, as can be seen directly from Eq.~(\ref{sxy}):
all functions in the integrand, except for $\omega_c(\epsilon)$ are
even in $\epsilon$. For finite $\mu\ll T$ the Hall coefficient is
linear in the chemical potential and thus linear in carrier density,
as can be seen from Eq.~(\ref{nl}).
\[ 
R_H\propto\frac{\mu v^2}{ecT^3} = \frac{n}{n_*^2ec}, \quad
n_* \propto \frac{T^2}{v^2}.
\]
The numerical coefficient in the above expression depends in the
precise nature of impurities (see Sec.~\ref{mft}).

\subsubsection{Single-gate setup}

At the time of writing, there is only one published report of a
Coulomb drag measurement in graphene-based double-layer system
\cite{tut}. In this experiment there is only one gate controlling
the carrier density in both layers. The carrier densities can then 
be found by solving two electro-static equations \cite{tut,csn}
\begin{subequations}
\label{est}
\begin{equation}
\label{nbl}
%eV_{BG} = e^2\frac{n_B + n_T}{C_1} + \mu_B,
eV_{BG} = \mu_B + e^2(n_B + n_T)/C_1,
\end{equation}
\begin{equation}
%\mu_B = e^2\frac{n_T}{C_2} + \mu_T,
\mu_B = \mu_T + e^2n_T/C_2,
\end{equation}
\end{subequations}
where $V_{BG}$ is the voltage applied to the bottom gate, $\mu_B$ and
$\mu_T$ are the chemical potentials of the bottom and top layer
respectively, $C_1$ is the capacitance of the oxide layer between the
gate and the bottom layer, and $C_2$ is the capacitance of the
inter-layer spacing.

Eqs.~(\ref{est}) were used in Ref.~\onlinecite{csn} to deduce that the
gate voltage $V_{BG}$ is proportional to the carrier density. Let us
estimate the density $n_B^*$ for which electrical and chemical
potentials of the bottom layer become comparable:
\begin{equation}
\label{nbs}
%\frac{n_B^*}{C_1} 
n_B^*/C_1
= v \sqrt{\pi n_B^*} \;\; \Rightarrow \;\;
n_B^* = \pi v^2 C_1 \sim 10^8 cm^{-2}.
\end{equation}
Here the numerical value is estimated using the parameters of the
experimental device \cite{tut,csn}. Thus the linear relation
$V_{BG}\propto n_B$ is valid for $n_B\gg n_B^*$, which is satisfied
for all densities considered in Ref.~\onlinecite{csn}.

In the vicinity of the Dirac point and in the presence of disorder
[see Eq.~(\ref{ndis})] all terms in Eqs.~(\ref{est}) are linear in
carrier density. As a result, both $n_B$ and $n_T$ are proportional to
the gate voltage $V_{BG}$ and in particular
\[
V_{BG} \propto \mu_B.
\]
According to Ref.~\onlinecite{tut} the carrier density in the top
layer depends on $V_{BG}$ only weakly and remains finite when the
bottom layer is tuned to the vicinity of the Dirac point, such that
$\mu_B<\mu_T$.  In such conditions the drag coefficient is described
by Eq.~(\ref{rd-12}). Since in this regime the gate voltage seems to
affect mostly the bottom gate of the device used in
Ref.~\onlinecite{tut}, we conclude that when the bottom layer is tuned
towards the Dirac point, the drag should vanish linearly with the gate
voltage. This conclusion is consistent with the experimental results
of Ref.~\onlinecite{tut}.

At the large carrier densities the drag should vanish as some power of
the gate voltage as shown in Fig.~\ref{r0}. Since in the experiment
the Fermi wavelength and the inter-layer spacing are of the same order
of magnitude, the decay of the drag coefficient at large $\mu$ is
described by Eq.~(\ref{rdlt-2}). Neglecting the weak dependence of
$n_T$ on the gate voltage reported in Ref.~\onlinecite{tut}, we
conclude that
\[
\rho_D\sim n_B^{-1},
\]
which qualitatively agrees with the experimental results.

Given that the drag vanishes both at small and large $\mu$ (or gate
voltages, or carrier densities), there must be a maximum at some
intermediate value of $\mu$, which is determined by temperature and
sample geometry. Thus the results of the perturbation theory, as shown
in Fig.~\ref{r0}, qualitatively describe all features of the drag
observed in the experiment. Moreover, for relatively small values of
the interaction parameter\cite{csn} $\alpha\approx 0.2$  our theory
yields a reasonable quantitative description of the effect. The
extension of our work for even stronger interaction and/or vanishing
disorder will be published separately\cite{us2}.

\subsubsection{Symmetric setup}

Another possibility \cite{ge2} is to align the Dirac points in the two
layers using a combination of gates and then apply a voltage ${\cal
  V}$ between the two layers, inducing same number of electrons in one
layer and holes in another (such that $n_1=n_2=n$ and
$\mu_1=-\mu_2=\mu$). In this case the results of Sec.~\ref{cdpt} and
\ref{fd} apply.

The inter-layer voltage ${\cal V}$ is related to the carrier density
by ${\cal V} = e^2 n/C + 2\mu$. The capacitance $C$ is the only
independently measurable coefficient in this relation and may be found
by measuring the Hall coefficient at a large chemical potential. Then
the ${\cal V}$-dependence of any quantity can be directly translated
into the density dependence (using Eq.~(\ref{nm}) to convert $\mu$ to
$n$).

\section{Conclusions}

We have presented the perturbative theory of the Coulomb
drag in ballistic graphene-based double-layer structures. Our theory
is applicable to the wide range of temperatures and carrier densities,
but is subject to the condition (\ref{att}). In addition we have
limited our discussion to the experimental \cite{tut} condition
(\ref{td}), the former is necessary to justify the theoretical
approach that we've adopted in this paper. As shown in
Sec.~\ref{al}, Eq.~(\ref{att}) allows us to simplify our
calculations by disregarding the effects of plasmon modes, Dirac
spectrum renormalization, and energy dependent impurity scattering
time. Eq.~(\ref{att}) also justifies our assumption that
impurity scattering dominates the transport properties in the system.

The main results of this paper can be summarized as follows.
Qualitatively, $\rho_D(\mu/T)$ has the same shape in all parameter
regimes: in the vicinity of the Dirac point $\rho_D\propto\mu^2/T^2$,
$\mu\sim T$ the drag reaches its maximum and then decays at $\mu\gg
T$. This decay occurs over a wide region of $\mu$ where $\rho_D$
cannot be described by a single power law. In particular, we have
analyzed three regimes: (i) in the limit $\alpha\rightarrow 0$ and
$d\rightarrow 0$ the drag coefficient is given by Eqs.~(\ref{rd-12s}),
(\ref{rd-12}), and (\ref{rdlt-2}), see also Table~\ref{table}; (ii)
for $\alpha\rightarrow 0$, but finite $d$ the drag coefficient
acquires logarithmic corrections, see Fig.~\ref{sk1}; (iii) for
intermediate interaction strength and $\mu\gg T$ we describe the
crossover between the logarithmic and the Fermi-liquid behavior, see
Eq.~(\ref{rf0}).  The latter occurs only at the largest values of
$\mu$, such that $\varkappa d\gg 1$. Thus our theory describes all
qualitative features observed in Ref.~\onlinecite{tut}.

Formally our results are applicable in the limit of weak
interaction. The actual value of $\alpha$ in physical graphene is
still the subject of a debate. Recent experiments
\cite{al1,al2} suggest that at experimentally relevant temperatures
the effective (or renormalized) interaction
parameter is rather small. In addition, if one takes into account
dielectric properties of the substrate and/or the insulating layer
between the two graphene sheets in the double-layer device \cite{csn},
then the effective value of $\alpha$ will be even smaller. 

For ultra-clean graphene, where transport is dominated by
electron-electron interaction, our theory should be generalized
for stronger interaction. In this case also the single-layer
conductivity becomes non-trivial. In our opinion, the
most adequate method for such calculations is the method of the
kinetic equation \cite{kas,kin,po2}. Our work in this direction will
be reported elsewhere \cite{us2}.

\begin{acknowledgments}

We acknowledge helpful conversations with A.K. Geim, M.I Katsnelson,
K.S. Novoselov, L. Ponomarenko, M. Sch\"utt, and A. Shnirman. This
research was supported by the Center for Functional Nanostructures of
the Deutsche Forschungsgemeinschaft (DFG) and by SPP 1459 ``Graphene''
of the DFG. M.T. is grateful to KIT for hospitality.

\end{acknowledgments}

\appendix

\section{Non-linear susceptibility in graphene}
\label{gamma}

Here we derive the non-linear susceptibility (\ref{gamma0}) and
consider a few limiting cases.

The general definition of the non-linear susceptibility is given by
Eq.~(\ref{gamma-el}), which we repeat here for convenience:
\begin{equation}
{\bf I}=\int d\bs{r}_1\int d\bs{r}_2
\bs{\Gamma}(\omega; \bs{r}_1,\bs{r}_2 ) 
V(\bs{r}_1)V(\bs{r}_2).
\label{gamma-el-A}
\end {equation}
\noindent
Furthermore, Eqs.~(\ref{nls}) and (\ref{g}) express the non-linear
susceptibility of a disordered conductor in terms of exact Green's
functions of the system (for detailed derivation see
Refs.~\onlinecite{kor,gam}):
\begin{eqnarray}
&&
{\bf \Gamma} = 
\int\frac{d\epsilon}{2\pi}
\Bigg[
      \left( \tanh\frac{\epsilon-\mu}{2T}
             - \tanh\frac{\epsilon+\omega-\mu}{2T}
      \right)
      \bs{\gamma}_{12}(\epsilon; \omega)
   %   \mbox{\boldmath$\gamma$}(\epsilon; \omega)
\nonumber\\
&&
\nonumber\\
&&
      +
      \left( \tanh\frac{\epsilon-\mu}{2T}
             - \tanh\frac{\epsilon-\omega-\mu}{2T}
      \right)
      \bs{\gamma}_{21}(\epsilon; -\omega)
\Bigg],
\label{nls-A}
\end{eqnarray}
\begin{equation}
\label{g-A}
\bs{\gamma}_{12}(\epsilon; \omega) = 
\Big[
      G^R_{12}(\epsilon + \omega) - G^A_{12}(\epsilon + \omega)
\Big]
G^{R}_{23}\left(\epsilon\right) 
{\bf \hat J}_3 
G^{A}_{31}\left(\epsilon\right).
\end{equation}
In contrast to the usual Fermi Liquid calculation
\cite{kor} we shift the chemical potential from the Green's 
functions into the distribution functions. 

Averaging over disorder restores translational invariance. Moreover,
in ballistic regime the Green's functions can be averaged
independently. However, in graphene the eigenfunctions of the Dirac
Hamiltonian are not plane waves. Focusing on a given valley and spin
projection, we can write down the Dirac Hamiltonian as
\begin{equation}
\label{dh}
{\cal H} = vk
\begin{pmatrix}
0 & e^{i\varphi_{\bs{k}}} \cr
e^{-i\varphi_{\bs{k}}} & 0
\end{pmatrix}, \;
\cos\varphi_{\bs{k}} = \frac{k_x}{k}, \;
\sin\varphi_{\bs{k}} = \frac{k_y}{k}.
\end{equation}
The electron field operator in the basis of eigenstates can be written
as 
\begin{equation}
\label{es}
\widehat \Psi (\bs{r}) 
=
\frac{1}{\sqrt{2}}\sum\limits_{\bs{k}, \nu}e^{i\bs{k}\bs{r}}
\begin{pmatrix}
\nu e^{-i\varphi_{\bs{k}}} \cr 1
\end{pmatrix}
\widehat c_{\bs{k}, \nu},
\end{equation}
where $\nu=\pm$ is the band index and the spinor [as well as the
Hamiltonian (\ref{dh})] is written in the sublattice space. In the
basis of the eigenstates the (disorder-averaged) Green's functions are
diagonal:
\begin{equation}
\label{gf0}
G_\nu^R(\epsilon, \bs{k}) =  
\frac{1}{\epsilon - E_\nu(\bs{k}) + i/2\tau(\bs{k})},
\quad E_\nu(\bs{k})=\nu v k.
\end{equation}
The impurity scattering time $\tau(\bs{k})$ may in general depend on
momentum of the scattering states (see Appendix~\ref{kin_ur}).

Now we can write the triangular vertex $\bs{\gamma}$ in the form:
\begin{eqnarray}
\label{gv-A}
&&
\bs{\gamma}(\epsilon; \omega, \bs{q}) = N \sum_{\nu\nu'}
\int\frac{d^2k}{(2\pi)^2} 
\left| \lambda^{\nu, \nu'}_{\bs{k}, \bs{k}+\bs{q}} \right|^2
\\
&&
\nonumber\\
&&
\quad\quad\quad
\times
{\rm Im}
G^R_{\nu'}(\epsilon + \omega; \bs{k}+\bs{q}) 
G^{R}_{\nu}\left(\epsilon; \bs{k}\right) 
\widehat{\bs{j}}^{\; \rm tr}_\nu 
G^{A}_{\nu}\left(\epsilon; \bs{k}\right).
\nonumber
\end{eqnarray}
The factor $N=4$ reflects the spin and valley degeneracy, and
\begin{equation}
\label{l-A}
\left|\lambda^{\nu,\nu'}_{\bs{k},\bs{k}'}\right|^2
=\frac{1}{2}
\left(1+\nu \nu' \frac{\bs{k}\bs{k}'}{kk'}\right).
\end{equation}
\noindent
The current operator \cite{and} should also be written in the basis of
the eigenstates where, unlike the original Bloch basis\cite{fal}, the
current operator depends on the direction of the quasi-particle
momentum
\begin{equation}
\label{jv-A}
\widehat{\bs{j}}^{\; \rm tr}_\nu 
= 2ev \nu \bs{n}_{\bs{k}}, 
\quad \bs{n}_{\bs{k}} = \bs{k}/k.
\end{equation}
\noindent
The factor of $2$ in Eq.~(\ref{jv-A}) appears due to the absence of
backscattering in graphene: the transport time $\tau_{\rm tr}$ is
twice the scattering time \cite{fal}.

Finally, in the ballistic regime $1/\tau\rightarrow 0$ and the
Green's functions (\ref{gf0}) can be written in the form 
\begin{subequations}
\label{gdel}
\begin{equation}
\label{grga}
G^R_{\nu} (\epsilon; \bs{k}) 
G^A_{\nu} (\epsilon; \bs{k}) \approx 
2\pi \tau(\epsilon) \delta
\Big( \epsilon - E_{\nu}(\bs{k}) \Big),
\end{equation}
\begin{equation}
\label{imgr}
{\rm Im} G^R_{\nu} (\epsilon; \bs{k}) 
\approx -\pi
\delta\Big( \epsilon - E_{\nu}(\bs{k}) \Big).
\end{equation}
\end{subequations}
Here we have replaced the momentum dependence of the scattering time
by the energy dependence given the $\delta$-function approximation to
the Green's functions. Note, that since Eq.~(\ref{gv-A}) does
not contain any energy integration, this dependence plays no role in
the triangular vertex $\bs{\gamma}$, which in ballistic regime takes
the form
\begin{eqnarray}
\label{gv-1}
&&
\bs{\gamma}(\epsilon; \omega, \bs{q}) = 
-N ev\tau(\epsilon)\sum_{\nu\nu'}\nu
\int d^2k \; \bs{n}_{\bs{k}}
\left| \lambda^{\nu, \nu'}_{\bs{k}, \bs{k}+\bs{q}} \right|^2
\\
&&
\nonumber\\
&&
\quad\quad\quad\quad\quad\quad\quad\quad
\times
\delta(\epsilon-\nu v k)
\delta\left(\epsilon+\omega-\nu' v|\bs{k}+\bs{q}|\right).
\nonumber
\end{eqnarray}
Using the $\delta$-functions in Eq.~(\ref{gv-1}) we notice, that in 
Eq.~(\ref{gv-1}) the momenta satisfy
\begin{equation}
\label{mom}
k^2 = \epsilon^2/v^2, \quad
\left(\bs{k}+\bs{q}\right)^2 = (\epsilon+\omega)^2/v^2,
\end{equation}
and therefore
\begin{equation}
\label{kq}
\bs{k}\bs{q} = \frac{(\epsilon+\omega)^2 - \epsilon^2 - v^2q^2}{2v^2}
=\frac{\omega^2 +2\epsilon\omega - v^2q^2}{2v^2}.
\end{equation}
Now we can replace the momentum dependence of the vertices
$\lambda^{\nu, \nu'}_{\bs{k}, \bs{k}+\bs{q}}$ by the frequency
dependence:
\begin{equation}
\label{l-1}
\left|\lambda^{\nu,\nu'}_{\bs{k},\bs{k}+\bs{q}}\right|^2
= 1 + \frac{\omega^2- v^2q^2}{4\epsilon(\epsilon+\omega)}.
\end{equation}
Then the triangular vertex (\ref{gv-1}) becomes
\begin{equation}
\label{gv-2}
\bs{\gamma}(\epsilon; \omega, \bs{q}) = -N ev\tau(\epsilon)
\left[ 1 + \frac{\omega^2- v^2q^2}{4\epsilon(\epsilon+\omega)} \right]
\bs{g},
\end{equation}
%where
\begin{equation}
\label{gl}
\bs{g} = \sum_{\nu\nu'}\nu
\int d^2k \; \bs{n}_{\bs{k}}
\delta(\epsilon-\nu v k)
\delta\left(\epsilon+\omega-\nu' v|\bs{k}+\bs{q}|\right).
\end{equation}
Clearly, the direction of the vector $\bs{g}$ coincides with the 
direction of $\bs{q}$:
\[
\bs{g} = A \bs{q} \quad\Rightarrow\quad
\bs{g} = \bs{q} \frac{(\bs{g}\bs{q})}{q^2},
\]
%where
\[
\bs{g}\bs{q} = \sum_{\nu\nu'}\nu
\int d^2k \frac{\bs{k}\bs{q}}{k}
\delta(\epsilon-\nu v k)
\delta\left(\epsilon+\omega-\nu' v|\bs{k}+\bs{q}|\right).
\]
Here we can again use Eq.~(\ref{kq}) and therefore
\begin{equation}
\label{gl-1}
\bs{g} = \bs{q} \frac{\omega^2 +2\epsilon\omega - v^2q^2}{2v^2q^2}
g_0,
\end{equation}
%where
\begin{equation}
\label{g0}
g_0 = \sum_{\nu\nu'}\nu \int \frac{d^2k}{k} \;
\delta(\epsilon-\nu v k)
\delta\left(\epsilon+\omega-\nu' v|\bs{k}+\bs{q}|\right).
\end{equation}
The remaining integration is straightforward and we find
\begin{equation}
\label{g0-r}
g_0 = \frac{4|\epsilon+\omega|\theta_0(\epsilon, \omega, q)}{
\sqrt{\left(v^2q^2-\omega^2\right)
\left(\left[2\epsilon+\omega\right]^2-v^2q^2\right)}}, 
\end{equation}
where ($\theta(x)$ is the Heaviside $\theta$-function)
\begin{widetext}
\begin{eqnarray}
\label{theta}
%&&
\theta_0(\epsilon, \omega, q) =
%\\
%&&
%\\
%&&
%\quad
%=
\theta(vq - |2\epsilon + \omega|)
\left[
\theta(-\omega-vq)-
\theta(\omega-vq)
\right]
%\\
%&&
%\\
%&&
%\quad\quad
+
\theta(vq-|\omega|)
\left[
\theta(2\epsilon + \omega - vq) -
\theta(- vq - 2\epsilon - \omega) \right].
\end{eqnarray}
Now the triangular vertex gamma can be written using Eqs.~(\ref{gv-2}), 
(\ref{gl-1}), (\ref{g0-r}), and (\ref{theta}) in the form (\ref{gamma0})
\begin{equation}
\label{gamma0-A}
\bs{\gamma}(\epsilon, \omega, q) = -\bs{q} N ev\tau(\epsilon)
\frac{\omega^2+2\epsilon\omega-v^2q^2}{2\epsilon v^2q^2}
\sqrt{\frac{\left[2\epsilon+\omega\right]^2-v^2q^2}{v^2q^2-\omega^2}}
\;
{\rm sgn}(\epsilon+\omega) \theta_0(\epsilon, \omega, q).
\end{equation}
The function $\theta_0(\epsilon, \omega, q)$ is antisymmetric under
the simultaneous change of sign of both frequencies
\[
\theta_0(\epsilon, \omega, q) = - \theta_0(-\epsilon, -\omega, q).
\]
Therefore the triangular vertex $\bs{\gamma}$ as a whole is symmetric
under the simultaneous change of sign of all variables:
\begin{equation}
\label{gs}
\bs{\gamma} (\epsilon, \omega, \bs{q}) =
\bs{\gamma} (-\epsilon, -\omega, -\bs{q}).
\end{equation}
Using this property in Eq.~(\ref{nls-A}) we find the expression for
the non-linear susceptibility in graphene:
\begin{subequations}
\label{nls-2}
\begin{equation}
\bs{\Gamma} = -  \frac{Nev\bs{q}}{2\pi} \;  g(\omega, q; \mu),
\end{equation}
\begin{equation}
g(\omega, q; \mu) = \int d\epsilon \tau(\epsilon) I(\epsilon, \omega)
F(\epsilon, \omega; q)\theta_0(\epsilon, \omega, q),
\end{equation}
\begin{equation}
F(\epsilon, \omega; q) = 
\frac{\omega^2+2\epsilon\omega-v^2q^2}{2\epsilon v^2q^2}
\sqrt{\frac{\left[2\epsilon+\omega\right]^2-v^2q^2}{v^2q^2-\omega^2}}
\;
{\rm sgn}(\epsilon+\omega),
\end{equation}
\begin{equation}
I(\epsilon, \omega) = \tanh\frac{\epsilon-\mu}{2T}
- \tanh\frac{\epsilon+\omega-\mu}{2T}
-\tanh\frac{\epsilon+\mu}{2T}
+ \tanh\frac{\epsilon+\omega+\mu}{2T}.
\end{equation}
\end{subequations}
\end{widetext}
The expression (\ref{gg}) then follows after a change of variables
indicated in the text preceding Eq.~(\ref{gg}) and explicitly
resolving the integration limits given by Eq.~(\ref{theta}).

Comparing Eq.~(\ref{nls-2}) with the standard Fermi liquid result of
Ref.~\onlinecite{kor}, we should recall that within the Fermi liquid
theory the Fermi energy is the largest energy scale in the problem.
In order to compare Eq.~(\ref{gs}) to the corresponding results of
Ref.~\onlinecite{kor}, we consider the limit
\[
\omega, vq \ll \epsilon \sim E_F.
\]
In this limit the leading contribution to the functions in the 
integrand in Eq.~(\ref{nls-2}) is given by
\[
F(\omega,vq\ll\epsilon) \approx 
\frac{\omega\epsilon}{v^2q^2}\frac{1}{\sqrt{v^2q^2-\omega^2}}
- \frac{\sqrt{v^2q^2-\omega^2}}{2v^2q^2},
\]

{
\begin{figure}
\begin{center}
\epsfig{file=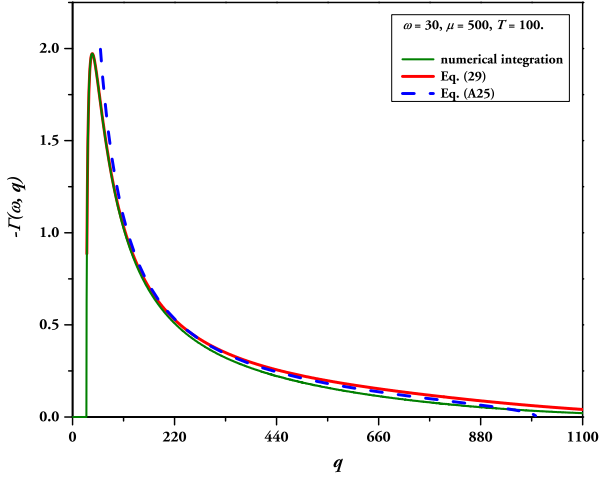,width=8cm}
\end{center}
\caption{[Color online] A numerical comparison of the approximations
  (\ref{ggr}) and (\ref{ggr-A2}). The former is shown by the red solid
  line, while the latter - by the blue dashed line. The solid green
  line is the numerical evaluation of Eq.~(\ref{nls-2}).}
\label{gamma_app}
\end{figure}
}

\[
I_2(\omega\ll\epsilon) \approx \frac{\omega}{2T}
\left[ \frac{\theta(\epsilon)}{\cosh^2\frac{\epsilon+\mu}{2T}}
-\frac{\theta(-\epsilon)}{\cosh^2\frac{\epsilon-\mu}{2T}} \right].
\]
Thus all three terms $F$, $I_2$, and $\theta_0$ in the integrand are
odd functions of $\epsilon$ and as a result the leading contribution
to non-linear susceptibility vanishes \cite{err,me1}. The subleading
contribution (stemming from the second term in the function $F$)
yields for $vq\gg\omega$
\begin{equation}
\label{ggr-A}
g(\omega, q) \approx - 4 \omega/vq,
\end{equation}
in agreement with Ref.~\onlinecite{kor}. Same result is given also by
Eq.~(\ref{ggr}) in the limit $\mu\gg T$ for $vq>\omega$.

The result (\ref{ggr-A}) was obtained assuming that typical values of
momentum $q$ are small compared to the Fermi momentum. This is
justified for large inter-layer spacing $\varkappa d\gg 1$, which was
assumed in Ref.~\onlinecite{kor}. However, in graphene-based samples
of Ref.~\onlinecite{tut} the inter-layer spacing is rather small and
$\varkappa d \sim 1$. In this case we can no longer assume the 
momentum $q$ to be small. Expanding the integrand in Eq.~(\ref{nls-2})
in the limit
%\[
$\omega\ll vq, \epsilon \sim E_F$,
%\]
we find
\[
F(\omega\ll vq,\epsilon) \approx -\frac{1}{2vq}
\sqrt{1-\frac{v^2q^2}{4\epsilon^2}},
\]
while the expansion of $I_2$ remains the same as above. Now the
frequency integration yields
\begin{equation}
\label{ggr-A2}
g(\omega\ll q) \approx -4 \frac{\omega}{vq} 
\sqrt{1-\frac{q^2}{4k_F^2}}.
\end{equation}
This expression approximates the non-linear susceptibility in the
region $\omega\ll vq < 2E_F$. Comparing Eq.~(\ref{ggr-A2}) to
Eq.~(\ref{ggr}) we note that both approximations work well in that
intermediate region, see Fig.~\ref{gamma_app}. At the same time,
Eq.~(\ref{ggr}) accounts better for the behavior at $q\sim\omega$
and also allows for momenta larger than $2k_F$. Note, that the
non-linear susceptibility (\ref{nls-2}) is real, the fact that
Eq.~(\ref{ggr-A2}) yields imaginary values for $q>2k_F$ is the
artifact of its approximate derivation.

\section{Kinetic equation approach to drag in graphene}
\label{kin_ur}

In this Appendix we derive the general expression for drag
conductivity $\sigma_D$ in the framework of the kinetic equation (this
is justified by requiring large single-layer conductivity
$\sigma_{1(2)}\gg e^2$, which is valid for $\mu\gg T$ or
$\mu\gg\tau$). By solving the two coupled equations for the
distribution functions of two graphene layers we will reproduce the
result (\ref{sd}) with the nonlinear susceptibility
$\bs{\Gamma}(\omega, \bs{q})$ given by Eqs.~(\ref{nls}) and
(\ref{gamma0}).

In Appendix~\ref{gamma} we have characterized the eigenstates
(\ref{es}) of the massless Dirac Hamiltonian $H = v \bs{\sigma}
\bs{k}$ [see also Eq.~(\ref{dh})] by the value of momentum $\bs{k}$
and the discrete variable $\nu = \pm 1$ indexing conduction and
valence bands. In this representation, the electron energy and
velocity are $E_\nu(\bs{k}) = \nu v k$ and $\bs{v} = \nu v
\bs{e}_{\bs{k}}$ (where $\bs{e}_{\bs{k}}=\bs{k}/k$ is the unit vector
pointing in the direction of momentum).

Now for the purposes of deriving the kinetic equation, we find it more
convenient to label the eigenstates by their energy $\epsilon$ and the
unit vector $\bs{e}_{\bs{v}}=\bs{v}/v$ . The particle momentum is then
$\bs{k} = \bs{e}_{\bs{v}} E_\nu(\bs{k})/v$ and the eigenstates are
normalized as follows
\begin{equation}
\int \frac{|\epsilon|\, d\epsilon\, d\bs{e}_{\bs{v}}}{(2\pi v)^2}\;
| \epsilon, \bs{e}_{\bs{v}} \rangle \langle \epsilon, \bs{e}_{\bs{v}} |
= 1.
\label{norm}
\end{equation} 

In the lowest order of the perturbation theory we neglect
electron-electron interaction within each layer and disregard the back
action of the drag current in the passive layer onto the distribution
function in the active layer. As a result the kinetic equation for the
active layer is effectively decoupled and has the form
\begin{equation}
\label{ke1}
e \bs{E}\bs{v}\, \frac{\partial f_a}{\partial\epsilon}
= \frac{\langle f_a \rangle - f_a}{\tau_a(\epsilon)}.
\end{equation}
Here $\bs{E}$ is the applied electric field and $\tau_a(\epsilon)$ is
the transport time due to disorder scattering. Index $a$ refers to the
active layer. The distribution function depends on $\epsilon$ and
$\bs{e}_{\bs{v}}$; angular brackets denote averaging with respect to
the velocity direction. Within linear response we substitute the
equilibrium distribution function $f_a^{(0)}$ in the left-hand side of
Eq.~(\ref{ke1}) and find the following solution
\begin{widetext}
\begin{equation}
 f_a
  = f_a^{(0)}
    -\tau_a(\epsilon) \frac{\partial f_a^{(0)}}{\partial\epsilon}\,
      e \bs{E}\bs{v}
  = f_a^{(0)}
    +\tau_a(\epsilon) f_a^{(0)} \big(1 - f_a^{(0)}\big) \frac{e \bs{E}\bs{v}}{T}.
 \label{fE}
\end{equation}
This is equation is written for the case of ``dirty'' graphene
$\tau_{dis}^{-1}\gg \tau_{ee}^{-1}$. The opposite limit of clean graphene will be
discussed in Ref.~\onlinecite{us2}. 

Consider now the passive layer. We denote the corresponding distribution
function $f_b$ and include the collision term describing inter-layer scattering.
The second kinetic equation has the form
\begin{equation}
\label{ke2}
 0
  = \frac{\langle f_b \rangle - f_b}{\tau_b}
    +\sum_{a,a',b'} w(a,b;\;a',b') \Big[
      f'_a f'_b (1 - f_a) (1 - f_b) - f_a f_b (1 - f_a') (1 - f_b')
    \Big].
\end{equation}
Here $w(a,b;\;a',b')$ is the probability of scattering $(a',b')
\mapsto (a,b)$, indices $a$, $b$, $a'$, and $b'$ label incoming and
scattered states in both layers.  Summation over these states is
carried out according to their normalization (\ref{norm}). 

The drag current can now be expressed as
\begin{equation}
 \bs{j}_D
  = e \sum_b \bs{v}_b f_b
  = e \sum_{a,a',b,b'} \tau_b \bs{v}_b\; w(a,b;\;a',b') \Big[
      f'_a f'_b (1 - f_a) (1 - f_b) - f_a f_b (1 - f'_a) (1 - f'_b)
    \Big].
\end{equation}
Now we substitute equilibrium distributions in layer $b$, and the
result (\ref{fE}) for $f_a$ and $f'_a$. Keeping only the terms linear
in the external field $\bs{E}$ and using the momentum conservation
law, we express the drag current as $j^\alpha_D =
\sigma^{\alpha\beta}_D E^\beta$. Using the time-reversal invariance of
the scattering probability, $w(a,b;\;a',b') = w(a',b';\;a,b)$, we
represent the resulting drag conductivity in the symmetric form
\begin{equation}
 \sigma^{\alpha\beta}_D
  = -\frac{e^2}{2T} \sum_{a,a',b,b'}
    (\tau'_b \bs{v}'_b - \tau_b \bs{v}_b)_\alpha
    (\tau'_a \bs{v}'_a - \tau_a \bs{v}_a)_\beta\;
    w(a,b;\;a',b')\; f'_a f'_b (1 - f_a) (1 - f_b).
 \label{sD}
\end{equation}
Here all distribution functions are taken at thermal equilibrium and the
superscripts are suppressed for brevity. Each of the four scattering times
entering the above equation is taken at the corresponding energy.

The transition probability $w(a,b;\;a',b')$ can be written with the
help of the Fermi golden rule:
\begin{equation}
 w(a,b;\;a',b')
  = \big| \langle a, b | U | a', b' \rangle \big|^2
    (2\pi)^3 \delta(\epsilon_a + \epsilon_b - \epsilon'_a - \epsilon'_b)\,
    \delta(\bs{k}_a + \bs{k}_b - \bs{k}'_a - \bs{k}'_b),
\end{equation}
where the matrix element of the inter-layer interaction includes the
Dirac factors (\ref{l-A}), re-written in terms of the velocities:
\begin{equation}
 \big| \langle a, b | U | a', b' \rangle \big|^2
  = \left|U\left(\bs{k}_a - \bs{k}'_a\right)\right|^2\,
    \frac{1 + \bs{e}_{\bs{v}}^{(a)}\bs{e}_{\bs{v}}^{(a')} }{2}\,
    \frac{1 + \bs{e}_{\bs{v}}^{(b)}\bs{e}_{\bs{v}}^{(b')}}{2}.
\end{equation}
With this matrix element, we can separate the quantities related to
layers $a$ and $b$ in the expression for drag conductivity
(\ref{sD}). This allows us to represent it in the form of
Eq.~(\ref{sd}):
\begin{equation}
 \sigma^{\alpha\beta}_D
  = \frac{e^2}{8T} \int \frac{d^2q\; d\omega\; |U(q)|^2}
      {(2\pi)^3\, \sinh^2\frac{\omega}{2T}}
    \Gamma^\alpha_a (\bs{q}, \omega) \Gamma^\beta_b(-\bs{q}, -\omega),
      \label{sDG} 
\end{equation}
\begin{equation}
\bs{\Gamma}_a(\bs{q}, \omega)
  = (e^{\omega/T}-1) \sum_{a,a'} f'_a(1 - f_a)
    (\tau_a' \bs{v}'_a - \tau_a \bs{v}_a)
    \frac{1 + \bs{e}_{\bs{v}}^{(a)} \bs{e}_{\bs{v}}^{(a')}}{2}\;
    (2\pi)^3 \delta(\epsilon_a - \epsilon'_a + \omega)\,
    \delta(\bs{k}_a - \bs{k}'_a + \bs{q}) 
\label{Gd}
\end{equation}
and the same formula for $\bs{\Gamma}_b$. Note the symmetry relations
$\bs{\Gamma}(-\bs{q}, -\omega) = -\bs{\Gamma}(\bs{q}, -\omega) =
\bs{\Gamma}(\bs{q}, \omega)$.

Let us now evaluate the expression (\ref{Gd}). Using the
energy-velocity basis and resolving the energy delta function, we
represent $\bs{\Gamma}$ as an integral over $\epsilon$ and over two
velocity directions $\bs{e}_{\bs{v}}$ and $\bs{e}_{\bs{v}}'$. With
equilibrium Fermi distribution functions, this yields
\begin{equation}
\bs{\Gamma}(\bs{q}, \omega)
  = \frac{\bs{q}}{8\pi v q^2} \int d\epsilon\,
    |\epsilon (\epsilon + \omega)|\,
    \left[
      \tanh\frac{\epsilon + \omega - \mu}{2T} - \tanh\frac{\epsilon - \mu}{2T}
    \right]
    J(\epsilon, \epsilon + \omega, q), 
\label{qG} 
\end{equation}
\begin{equation}
 J(\epsilon, \epsilon', q)
  = \int d\bs{e}_{\bs{v}}\, d\bs{e}_{\bs{v}}'\,
    (\tau' \bs{q} \bs{e}_{\bs{v}}' - \tau \bs{q} \bs{e}_{\bs{v}})
    (1 + \bs{e}_{\bs{v}} \bs{e}_{\bs{v}}')
    \delta(
      \epsilon \bs{e}_{\bs{v}} - \epsilon' \bs{e}_{\bs{v}}' + v \bs{q}
    ).
\end{equation}
The two-dimensional delta function in the latter integral fixes both
$\bs{e}_{\bs{v}}$ and $\bs{e}_{\bs{v}}'$. We substitute $\bs{q}$ from
the argument of the delta function into the rest of the integrand and then
average the delta function over directions of $\bs{q}$. After such an
averaging the integrand depends only on the angle $\phi$ between
$\bs{e}_{\bs{v}}$ and $\bs{e}_{\bs{v}}'$.
\begin{subequations}
\label{J}
%\begin{eqnarray}
%&&
\begin{equation}
 J(\epsilon, \epsilon', q)
  = \int \frac{d\phi}{v^2 q}\,
    [\tau' \epsilon' + \tau \epsilon
     -(\tau' \epsilon + \tau \epsilon')\cos\phi
    ]
    (1 + \cos\phi)
    \delta\left(
      \sqrt{\epsilon^2 + {\epsilon'}^2 - 2 \epsilon\epsilon' \cos\phi} - v q
    \right) 
%\nonumber\\
%&&
%\nonumber\\
%&&
%\quad\quad\quad\quad
= \tilde J(\epsilon, \epsilon', q) + \tilde J(\epsilon', \epsilon, q),
%\end{eqnarray}
\end{equation}
\begin{eqnarray}
\tilde J(\epsilon, \epsilon', q)  = \tau(\epsilon)
\frac{v^2 q^2 + \epsilon^2 - {\epsilon'}^2}{v \epsilon^2 \epsilon'}
\sqrt{
      \frac{(\epsilon + \epsilon')^2 - v^2 q^2}
           {v^2 q^2 - (\epsilon - \epsilon')^2}
    } . 
\end{eqnarray}
\end{subequations}
This result should be treated as zero if the argument of the square
root is negative [this fact was previously expressed in terms of the
additional factor $\theta_0(\epsilon, \omega, q)$]. 
Taking advantage of the symmetry of Eq.~(\ref{J}), we recast
$\bs{\Gamma}$ in the form of Eqs.~(\ref{nls}) and (\ref{gamma0}). In
particular, we identify the function (\ref{J}) with Eq.~(\ref{gamma0})
as
\[
\bs{\gamma}(\epsilon, \omega, q) = - N\frac{\bs{q}}{2q^2} \;
|\epsilon (\epsilon + \omega)|
\tilde J(\epsilon, \epsilon+\omega, q), 
\]
where we have multiplied the result by $N$ to account for the spin and
valley degeneracy. Note, that in this Appendix $\tau$ stands for the
transport scattering time, unlike Eq.~(\ref{gamma0}), where $\tau$ is
just the mean free time. In graphene these two quantities differ by a
factor of two \cite{fal}.

%\end{widetext}

\section{Polarization operator in graphene}
\label{pol_op}

In the basis of exact eigenstates we can use the standard expression
for the polarization operator, including the vertices
$\lambda^{\nu,\nu'}_{{\bf k},{\bf k}'}$ and summing over the two
bands:
%\begin{widetext}
\begin{equation}
\label{po}
\Pi^R(\omega, \bs{q}) = -4 \sum_{\nu\nu'}
\int\frac{d^2k}{(2\pi)^2} 
| \lambda^{\nu,\nu'}_{\bs{k},\bs{k}+\bs{q}}|^2
\frac{n_F(\bs{k}) - n_F({\bs{k}+\bs{q}})}
%\right)| \lambda^{\alpha,\alpha'}_{{\bf k},{\bf k}+{\bf q}}|^2}
{\omega - E_\nu(\bs{k}+\bs{q}) + E_{\nu'}(\bs{k}) + i\eta}
\end{equation}
[the prefactor of $4$ is due to spin and valley degeneracy; the
overall sign is chosen in such a way that the static polarization
operator at $q=0$ yields the density of states (\ref{tdos})].
Here $n_F(\bs{k})$ stands for the Fermi distribution. 

The polarization operator was calculated in detail in
Ref.~\onlinecite{po2}. Nevertheless, we will add some details in order
to make the paper self-contained. The complete expression for the
polarization operator might also be useful for numerical computations.

In order to simplify the expression for the polarization operator we
multiply Eq.~(\ref{po}) by the integral of a $\delta$-function, which
is unity:
%\begin{widetext}
\begin{equation}
\label{p1}
\Pi^R = -4 \sum_{\nu\nu'} \int\frac{d^2k}{(2\pi)^2} 
| \lambda^{\nu,\nu'}_{\bs{k},\bs{k}+\bs{q}}|^2
\frac{\left[n_F(\bs{k}) - n_F({\bs{k}+\bs{q}})\right]}
{\omega - E_\nu(\bs{k}+\bs{q}) + E_{\nu'}(\bs{k}) + i\eta}
\int d\epsilon_1 \delta\left(\epsilon_1-E_{\nu'}(\bs{k})\right)
\int d\epsilon_2 \delta\left(\epsilon_2-E_\nu(\bs{k}+\bs{q})\right).
\end{equation}
%\end{widetext}
Now we can use the $\delta$-functions to express the integrand in
Eq.~(\ref{po}) in terms of $\epsilon_i$. After that the momentum
integral will only contain the two $\delta$-functions and can be
evaluated analytically similarly to how it was done in
Appendix~\ref{gamma} for the non-linear susceptibility. Then
the polarization operator takes the form
\begin{equation}
\label{p2}
\Pi^R = -\int \frac{d\epsilon_1 d\epsilon_2}{\epsilon_1\epsilon_2}
[n_F(\epsilon_1) - n_F(\epsilon_2)]
\frac{(\epsilon_1+\epsilon_2)^2 - v^2q^2}
{\omega-\epsilon_2+\epsilon_1 +i\eta}
F(\epsilon_1,\epsilon_2),
\end{equation}
where
\begin{equation}
\label{f1}
F(\epsilon_1,\epsilon_2) = \sum_{\nu\nu'} 
\int\frac{d^2k}{(2\pi)^2}
\delta\left(\epsilon_1-\nu' v k\right)
\delta\left(\epsilon_2-\nu v |\bs{k}+\bs{q}| \right).
\end{equation}
The calculation of this function can be performed along the lines of
Appendix~\ref{gamma}. The result is
\begin{equation}
\label{f2}
F(\epsilon_1,\epsilon_2) = \frac{\epsilon_1|\epsilon_2|}{\pi^2v^2}
\frac{1}
{\sqrt{\left[(\epsilon_1+vq)^2-\epsilon_2^2\right]
\left[\epsilon_2^2-(\epsilon_1-vq)^2\right]}} \;
\Theta(\epsilon_1,\epsilon_2),
\end{equation}
where
\begin{eqnarray}
\label{th-p}
\Theta(\epsilon_1,\epsilon_2) =
\theta(\epsilon_1>0)
\theta\left[(\epsilon_1-vq)^2<\epsilon_2^2
<(\epsilon_1+vq)^2\right]
-
\theta(\epsilon_1<0)
\theta\left[(\epsilon_1+vq)^2<\epsilon_2^2
<(\epsilon_1-vq)^2\right].
\end{eqnarray}
The $\theta$-functions are the result of imposing the condition that
the cosine of the angle between the two momenta in Eq.~(\ref{kq}) is
less than unity. In other words, the expression under the square root
in Eq.~(\ref{f2}) has to be positive (and thus
$F(\epsilon_1,\epsilon_2)$ is a real function).  

The resulting expression can be simplified by the series of simple
transformations: (i) change the sign of $\epsilon_i$ in the second term
in Eq.~(\ref{th-p}); (ii) resolve the $\theta$-functions in order to 
identify the integration limits; (iii) introduce the sum and difference 
\[
z_1 = \epsilon_1+\epsilon_2, \quad
z_2 = \epsilon_1-\epsilon_2,
\]
(iv) introduce the dimensionless variables 
\[
Q=\frac{q}{2T}, \quad W=\frac{\omega}{2T}, \quad x=\frac{\mu}{T}. 
\]
As a result we arrive at the following expression
\begin{eqnarray}
\label{pgt}
&&
\Pi^R = \frac{q}{4\pi^2 v} \int\limits_0^1 
\int\limits_0^1\frac{dz_1dz_2}{z_1\sqrt{(1-z_1^2)(1-z_2^2)}}
\left[
(z_1^{-2}-1)\left(\frac{Q}{z_2 Q + W + i\eta} + 
\frac{Q}{z_2 Q - W - i\eta} \right) J_1(z_1^{-1}, z_2, x_i) \right.
\\
&&
\nonumber\\
&&
\quad\quad\quad\quad\quad\quad\quad
\quad\quad\quad\quad\quad\quad\quad
\quad\quad\quad\quad\quad\quad\quad
+
\left.
(1-z_2^2)\left(\frac{Q}{z_1^{-1} Q + W + i\eta} + 
\frac{Q}{z_1^{-1} Q - W - i\eta} \right) J_2(z_1^{-1}, z_2, x_i)
\right]
\nonumber
\end{eqnarray}
\begin{equation}
J_{1(2)}(z_1, z_2, x) = \tanh\frac{(z_1+z_2)Q+x}{2} +
\tanh\frac{(z_1+z_2)Q-x}{2}
\mp
\tanh\frac{(z_1-z_2)Q+x}{2}
\mp
\tanh\frac{(z_1-z_2)Q-x}{2}
\end{equation}
In particular, it is instructive to further simplify the imaginary
part of the polarization operator:
\begin{subequations}
\label{impg}

\begin{equation}
{\rm Im}\Pi^R(\omega, \bs{q}) = \frac{q}{4 \pi v}
\left[\theta(|W|>Q) P_1(W, Q)+
\theta(|W|<Q) P_2(W, Q)\right];
\end{equation}

\begin{equation}
\label{imp1}
P_1(W, Q) = 
\frac{{\rm sgn}W}{\sqrt{W^2/Q^2-1}}
\int\limits_0^1dz \sqrt{1-z^2} \;
I_1(z; Q, W, x);
\end{equation}

\begin{equation}
\label{imp2}
P_2(W, Q) = \theta(|W|<Q)
\frac{{\rm sgn}W}{\sqrt{1-W^2/Q^2}}
\int\limits_1^\infty dz \sqrt{z^2-1} \;
I_1(z; Q, W, x);
\end{equation}

\begin{eqnarray}
\label{i1}
I_1(z; Q, W, x) =
\tanh\frac{zQ+W+x}{2} +
\tanh\frac{zQ+W-x}{2} -
\tanh\frac{zQ-W-x}{2} -
\tanh\frac{zQ-W+x}{2}.
\end{eqnarray}

\end{subequations}

Comparing Eqs.~(\ref{impg}) and (\ref{gg}) we conclude, that despite
clear similarity, these expression are {\it not} proportional to each
other. Therefore the proportionality between the non-linear
susceptibility and the imaginary part of the polarization operator
mentioned in Ref.~\onlinecite{kor} is not a general theorem, but
rather a property of the limiting cases considered in
Ref.~\onlinecite{kor}.

Having the full expression for the polarization operator it is
straightforward to derive the well-known expressions:
\begin{equation}
\label{ptl1}
\Pi^R(\mu=\omega=0; T\ll vq) \approx q/4v, 
\end{equation}
\begin{eqnarray}
\label{pgd3}
\Pi^R(\mu=\omega=0; T\gg vq) \approx 4 T \ln 2 /(\pi v^2) ,
\end{eqnarray}
\begin{equation}
\label{pgfls}
\Pi^R(q\ll k_F; \omega=T=0)\approx 2k_F/\pi v,
\end{equation}
\begin{equation}
\label{pgflr}
{\rm Re}  \Pi^R(q\ll k_F; T=0) \approx \frac{2k_F}{\pi v}
\left[ 1 -
\frac{|\omega|\theta(\omega^2>v^2q^2)}{\sqrt{|\omega|-vq}} \; 
\right],
\quad\quad
%\end{equation}
%\begin{equation}
%\label{pgfli}
{\rm Im}  \Pi^R(q\ll k_F; T=0) \approx\frac{2k_F}{\pi v}
\frac{\omega \; \theta(|\omega|<vq)}{\sqrt{v^2q^2-\omega^2}} \; .
\end{equation}

%\end{widetext}

%\newpage

%\begin{widetext}

\section{Numerical evaluation of the drag coefficient}
\label{numres}

In this Appendix we show the results of the numerical evaluation of
the drag coefficient using Eqs.~(\ref{d1}), (\ref{sd}), and
(\ref{s0}).  The interaction propagator (\ref{d12}) was calculated
using the polarization operator calculated in Appendix~\ref{pol_op} in
the absence of disorder, see Eq.~ (\ref{pgt}). The non-linear
susceptibility was evaluated using Eqs.~(\ref{gg}). The particle
density was found from Eq.~(\ref{nm}). The particular values of the
inter-layer spacing $d$, temperature $T$, and the interaction
parameter $\alpha$ were chosen to resemble possible realizations of
the drag measurement in graphene-based devices \cite{tut,ge2}.

\begin{figure}[h]
\vspace{7pt}
\begin{center}
\includegraphics[height=150pt]{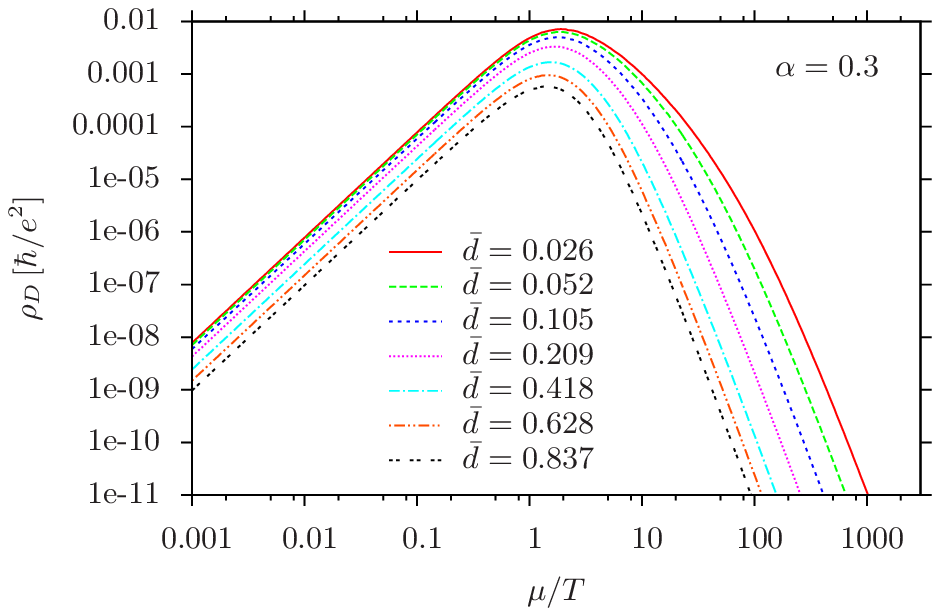}\hspace*{25pt}
\includegraphics[height=150pt]{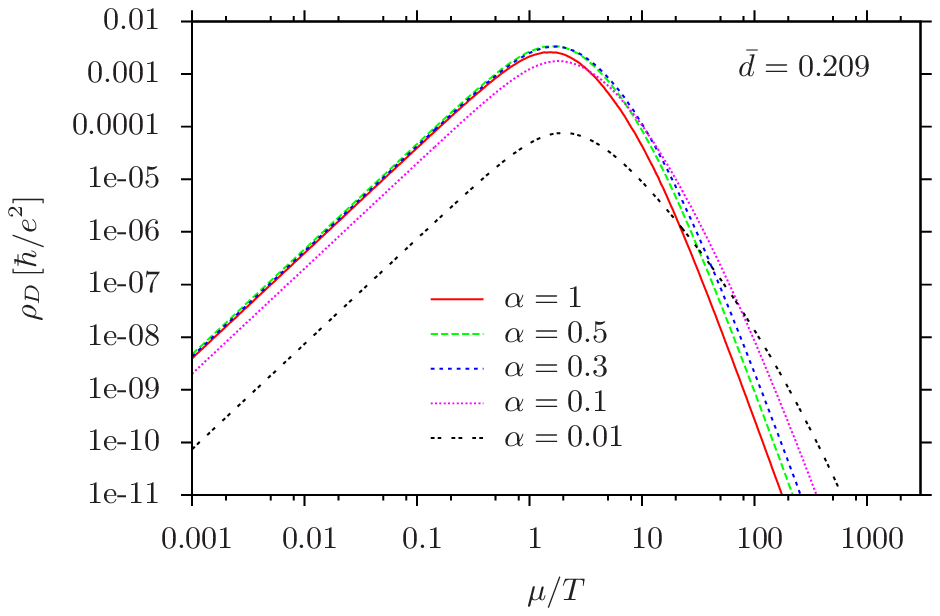}
\end{center}
\caption{Numerical estimate of the drag resistance. The left panel
  shows $\rho_D(\mu/T)$ for $\alpha=0.3$ (see Ref.\onlinecite{csn})
  and various values of $d$ (see Table~\ref{tab}). The right panel
  shows $\rho_D(\mu/T)$ for $\bar{d}=2 T d/ v$ and various values of
  $\alpha$.}
\label{fig:LOG}
\end{figure}

\begin{figure}[h]
\begin{center}
\includegraphics[height=160pt]{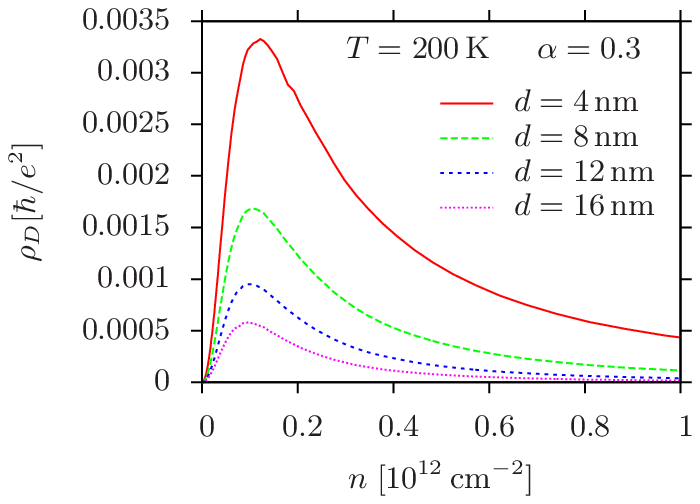}\hspace*{25pt}
\includegraphics[height=160pt]{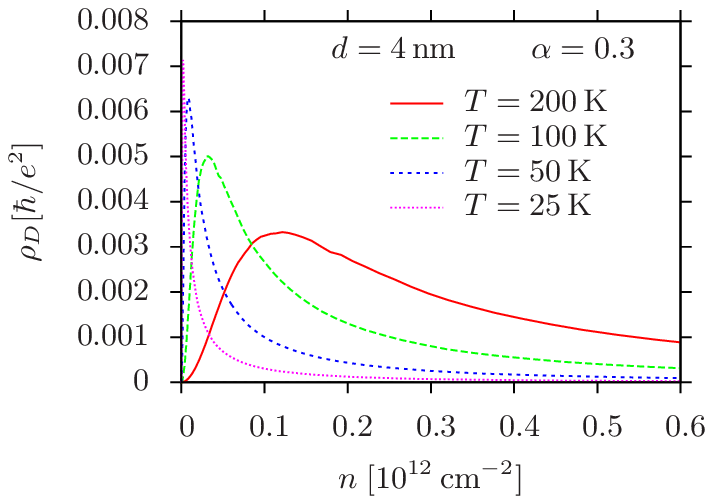}\\
\end{center}
\caption{The data from Fig.~\ref{fig:LOG} in the linear
  scale. The chemical potential is converted to the carrier density,
  $n$, using Eq.~(\ref{nm}).}
\label{fig:AB}
\end{figure}

\begin{table}[h]
\caption{Values of the dimensionless parameter $\bar{d}=2 T d/\hbar v$
  for different inter-layer separations $d$ and temperatures $T$.}
\begin{center}
\begin{ruledtabular}
%\begin{tabular}{p{0.04\textwidth} p{0.08\textwidth} || p{0.1\textwidth} | p{0.1\textwidth} | p{0.1\textwidth} |  p{0.1\textwidth} | p{0.1\textwidth} | p{0.1\textwidth} } 
\begin{tabular}{ccccccc}
$T$ & $d=4$\,nm & 6\,nm & 8\,nm & 12\,nm & 16\, nm & 18\,nm \\ 
\noalign{\smallskip}
\hline \noalign{\smallskip}
25\,K & 0.026 & 0.039 & 0.052 & 0.078 & 0.105 & 0.118 \\ \noalign{\smallskip}
50\,K & 0.052 & 0.078 & 0.105 & 0.157 & 0.209 & 0.235  \\  \noalign{\smallskip}
100\,K & 0.105 & 0.157 & 0.209 & 0.314  & 0.418 & 0.471  \\ \noalign{\smallskip}
200\,K & 0.209 & 0.314 & 0.418 & 0.628 & 0.837 & 0.942  
\end{tabular}
\end{ruledtabular}
\end{center}
\label{tab}
\end{table}

\newpage

\end{widetext}

The maximal values of $\rho_D$ are apparently reached for $\alpha \sim
0.5$. The drag resistance at the peak is a non-monotonous function of
$\alpha$, since we are calculating the drag conductivity within the
lowest-order perturbation theory, but still keep $\alpha\ne 0$ in the
denominator of the interaction propagator in order to describe
screening effects. The peak values of $\rho_D$ are achieved for
carrier densities such that $\mu\sim T$ (only weakly
depending on $d$).


\begin{thebibliography}{99}
\bibitem{fn3} Expressing the electron-hole asymmetry as a derivative
  of the single-layer conductivity with respect to the chemical
  potential \cite{dsm}, one might conclude \cite{me1} that for the
  case of short-ranged disorder (where $\sigma_i$ is independent of
  $\mu_i$ or carrier density \cite{fal}) the drag should
  vanish. However, the non-linear susceptibility in ballistic regime
  contains another term, which is proportional\cite{fn2} to $\partial
  \tau/\partial\mu$. Hence there is no accidental cancellation of the
  drag and we can conclude that taking into account energy dependence
  of the scattering time does not lead to qualitative changes in the
  theory.
\bibitem{fn4} For $\mu\gg T$
  the drag conductivity is largely determined by $vq\gg\omega\sim
  T\gg\tau^{-1}$. In turn, the non-linear susceptibility is determined
  by $\epsilon\sim T\gg\tau^{-1}$. Thus we can use the theory
  developed in Appendix~\ref{gamma}, even though technically
  we can only represent the Green's function by Eq.~(\ref{gdel}) for
  $\epsilon\gg\tau^{-1}$. Moreover, in the region $vq,
  \omega\ll\tau^{-1}$ we should take into account multiple impurity
  scattering, i.e. dress the non-linear susceptibility by
  diffusons. In the ballistic regime, however, the region $vq,
  \omega\ll\tau^{-1}$ is irrelevant.
\end{thebibliography}
\end{document}